\documentclass[prd,eqsecnum,twocolumn,amsfonts,amssymb]{revtex4}

\usepackage{CJK}

\usepackage{graphicx}

\usepackage{bm}

\newcommand{\beq}{\begin{equation}}
\newcommand{\eeq}{\end{equation}}
\newcommand{\beqs}{\begin{eqnarray}}
\newcommand{\eeqs}{\end{eqnarray}}

\newcommand{\drawsquare}[2]{\hbox{%
\rule{#2pt}{#1pt}\hskip-#2pt
\rule{#1pt}{#2pt}\hskip-#1pt
\rule[#1pt]{#1pt}{#2pt}}\rule[#1pt]{#2pt}{#2pt}\hskip-#2pt
\rule{#2pt}{#1pt}}
\newcommand{\fund}{\raisebox{-.5pt}{\drawsquare{6.5}{0.4}}}
\newcommand{\sym}{\raisebox{-.5pt}{\drawsquare{6.5}{0.4}}\hskip-0.4pt%
        \raisebox{-.5pt}{\drawsquare{6.5}{0.4}}}
\newcommand{\asym}{\raisebox{-3.5pt}{\drawsquare{6.5}{0.4}}\hskip-6.9pt%
        \raisebox{3pt}{\drawsquare{6.5}{0.4}}}

\begin{document}

\begin{CJK*}{UTF8}{}

\title{Renormalization-Group Evolution and Nonperturbative Behavior of 
Chiral Gauge Theories with Fermions in Higher-Dimensional Representations}

\author{Yan-Liang Shi (\CJKfamily{bsmi}石炎亮) and Robert Shrock} 

\affiliation{C. N. Yang Institute for Theoretical Physics, 
Stony Brook University, Stony Brook, N. Y. 11794 }

\begin{abstract}

We study the ultraviolet to infrared evolution and nonperturbative behavior of
a simple set of asymptotically free chiral gauge theories with an SU($N$) gauge
group and an anomaly-free set of $n_{S_k}$ copies of chiral fermions
transforming as the symmetric rank-$k$ tensor representation, $S_k$, and
$n_{\bar A_\ell}$ copies of fermions transforming according to the conjugate
antisymmetric rank-$\ell$ tensor representation, $\bar A_\ell$, of this group
with $k, \ \ell \ge 2$. As part of our study, we prove a general theorem
guaranteeing that a low-energy effective theory resulting from the dynamical
breaking of an anomaly-free chiral gauge theory is also anomaly-free. We
analyze the theories with $k=\ell=2$ in detail and show that there are only a
finite number of these.  Depending on the specific theory, the ultraviolet to
infrared evolution may lead to a non-Abelian Coulomb phase, or may involve
confinement with massless composite fermions, or fermion condensation with
dynamical gauge and global symmetry breaking. We show that $S_k \bar A_k$ 
chiral gauge theories with $k \ge 3$ are not asymptotically free.  We
also analyze theories with fermions in $S_k$ and $\bar A_\ell$ representations
of SU($N$) with $k \ne \ell$ and $k, \ \ell \ge 2$.

\end{abstract}


\maketitle

\end{CJK*}


\section{Introduction}
\label{intro}

The properties of chiral gauge theories are of fundamental field-theoretic
interest.  We recall the definition of such theories: with the fermions written
in left-handed chiral form, they transform as complex representations of the
gauge group.  It is natural to restrict one's consideration to asymptotically
free chiral gauge theories since this guarantees that in the deep ultraviolet
(UV) at large Euclidean reference momentum scales $\mu$, the gauge coupling is
small and hence such theories are perturbatively calculable.  In order to avoid
a triangle anomaly in gauged currents that would spoil renormalizability, one
also requires that the sum of the contributions from the fermions to this
anomaly must vanish.  Given that the theory is asymptotically free, its running
gauge coupling $g(\mu)$ increases as the scale $\mu$ decreases toward the
infrared (IR).  A basic goal of quantum field theory is to understand this UV
to IR evolution of the theory.  A chiral gauge theory is said to be irreducibly
chiral if it does not contain any vectorlike subsector.  In this case, the
chiral gauge symmetry precludes any fermion mass terms in the underlying
Lagrangian.  We shall focus on such theories here.  If the fermion content of
the theory satisfies the 't Hooft global anomaly-matching conditions, then, as
the coupling becomes sufficiently strong in the infrared, the gauge interaction
may confine and produce massless gauge-singlet composite spin-1/2 fermions
\cite{thooft1979}-\cite{cgt2}.  Alternatively, at some scale $\mu = \Lambda$,
the gauge interaction could become strong enough to produce one or more
bilinear fermion condensates, thereby spontaneously breaking gauge and global
chiral symmetries \cite{dfcgt,cgt,cgt2},\cite{mac}-\cite{uvcomplete_etc}.  Just
as quarks gain constituent masses in quantum chromodynamics, the fermions
involved in these condensates gain dynamical masses of order $\Lambda$.  One
then integrates them out to construct an effective theory with the remaining
massless fields that is applicable as the reference scale $\mu$ decreases below
$\Lambda$. In general, there can be several such stages of condensate formation
and symmetry breaking, with an associated sequence of effective field theories
that describe the physics at different intervals of $\mu$.  A third type of
renormalization-group (RG) evolution that occurs in certain theories is that
although the running gauge coupling grows as $\mu$ decreases, it reaches an
infrared fixed point at sufficiently weak coupling that neither confinement nor
fermion condensation occurs, and instead there is a (deconfined) non-Abelian
Coulomb behavior in the infrared. Asymptotically free chiral gauge theories
have been explored in the past in efforts to understand some of the questions
that the Standard Model (SM) can accommodate but does not explain, such as the
origin of fermion generations and the spectrum of quark and lepton masses.
Some work along these lines includes \cite{thooft1979}-\cite{uvcomplete_etc}.

In this paper we continue the recent investigations in \cite{cgt,cgt2} of the
UV to IR evolution of chiral gauge theories.  In general, it is valuable to
examine as many such theories as possible to gain insight into their behavior.
Here we study a class of chiral gauge theories in which the underlying theory
in the deep UV has only fermions transforming according to higher-dimensional
representations of the gauge group, but no fermions in the fundamental or
conjugate fundamental representation, and is irreducibly chiral, with no
vectorlike subsector.  Specifically, we construct and study asymptotically free
chiral gauge theories with an SU($N$) gauge group and an anomaly-free set of
$n_{S_k}$ copies of chiral fermions transforming as the symmetric rank-$k$
tensor representation, denoted $S_k$, and $n_{\bar A_\ell}$ copies of fermions
transforming according to the conjugate antisymmetric rank-$\ell$ tensor
representation, denoted $\bar A_\ell$, of this group, with $k, \ \ell \ge 2$
\ \cite{sbara}.  We will equivalently refer to these copies as flavors.  One
basic property of these theories that differs with previously studied chiral
gauge theories may be highlighted at the outset, namely the property that the
combined requirements of asymptotic freedom and anomaly cancellation constrain
the models so strongly that at most only a finite number of $S_k \bar A_\ell$
models satisfy these requirements.  In contrast, previously studied chiral
gauge theories typically form infinite families, such as the family of models
with SU($N$) gauge groups and chiral fermions transforming as 
$S_k + (N+4) \bar F$ (defined for any $N \ge 3$), and the family with chiral 
fermions transforming as $A_k + (N-4) \bar F$ (defined for any $N \ge 5$), and
extensions thereof including vectorlike subsectors.  As part of our study, we
prove a general theorem guaranteeing that a low-energy effective theory
resulting from the dynamical breaking of an anomaly-free chiral gauge theory is
also anomaly-free. We carry out a detailed analysis of the simplest case,
$k=\ell=2$, i.e., $S_2 \bar A_2$ theories. We give a complete enumeration of
the finite set of such theories satisfying the constraints of anomaly
cancellation and asymptotic freedom, and investigate the interesting variety of
patterns of UV to IR evolution that these theories exhibit.  We then proceed to
analyze theories with higher values of $k$ and/or $\ell$.  We restrict our
consideration here to chiral gauge theories with only gauge and fermion fields,
but without any scalar fields.  

This paper is organized as follows.  In Sect. \ref{methods} we briefly review
our general theoretical framework and methods of analysis.  In
Sect. \ref{descendant_anomalyfree} we prove that dynamical breaking of an
anomaly-free chiral gauge theory yields a low-energy effective theory that is
also anomaly-free. In
Sect. \ref{sabarmodel} we present our new set of chiral gauge
theories with gauge group SU($N$) and chiral fermions transforming according to
(an anomaly-free set of) the rank-2 symmetric and conjugate antisymmetric
tensor representations of the gauge group. Sections \ref{su5} and
\ref{sabar_nge6} are devoted to detailed analyses of the UV to IR evolution 
of models of this type with an SU(5) gauge group and with an SU($N$) gauge
group with $N \ge 6$, respectively. In Sect. \ref{kge3} we show that there are
no asymptotically free $S_k \bar A_k$ chiral gauge theories with $k \ge 3$.  In
Sect. \ref{skabarell} we analyze $S_k \bar A_\ell$ chiral gauge theories with 
$k \ne \ell$ and $k, \ \ell \ge 2$.  Sections \ref{s3abar2} and \ref{s2abar3}
are devoted to the two simplest of these, namely the $S_3 \bar A_2$ and $S_2
\bar A_3$ theories.  We give our conclusions are  in Sect. \ref{conclusions}.
Some auxiliary formulas are included in two appendices. 


\section{Methods of Analysis}
\label{methods}

In this section we briefly review the methods of analysis that we use. As noted
above, we consider asymptotically free chiral gauge theories with gauge group
$G={\rm SU}(N)$ and denote the running gauge coupling measured at a Euclidean
momentum scale as $g(\mu)$.  It is also convenient to use the quantities
$\alpha(\mu) = g(\mu)^2/(4\pi)$ and $a(\mu) \equiv g(\mu)^2/16\pi^2$. We will
often suppress the argument $\mu$ in these running quantities. Without loss of
generality, we write all fermion fields in terms of left-handed chiral
components.

The ultraviolet to infrared evolution of the gauge coupling is described by the
beta function, $\beta_g = dg/dt$, or equivalently,
$\beta_\alpha = d\alpha/dt = [g/(2\pi)]\beta_g$, 
where $dt = d\ln \mu$. This has the series expansion
\beq
\beta_\alpha = -2\alpha \sum_{\ell=1}^\infty b_\ell \, a^\ell =
-2\alpha \sum_{\ell=1}^\infty \bar b_\ell \, \alpha^\ell \ ,
\label{beta}
\eeq
where we have extracted an overall minus sign, $b_\ell$ is the $\ell$-loop
coefficient, and it will be useful to define the reduced $\ell$-loop
coefficient, $\bar b_\ell = b_\ell/(4\pi)^\ell$.  The $n$-loop beta function,
denoted $\beta_{\alpha,n\ell}$, is given by Eq. (\ref{beta}) with the upper
limit on the $\ell$-loop summation equal to $n$ instead of $\infty$. The
requirement of asymptotic freedom means that $\beta_\alpha < 0$ for small
$\alpha$, which holds if $b_1 > 0$.  The one-loop and two-loop coefficients
$b_1$ \cite{b1} and $b_2$ \cite{b2} are independent of the scheme used for
regularization and renormalization, while the $b_\ell$ with $\ell \ge 3$ are
scheme-dependent.

With $b_1 > 0$, if $b_2 < 0$, then the two-loop beta function,
$\beta_{\alpha,2\ell}$, has an IR zero at $a_{_{IR,2\ell}}=-b_1/b_2$, or
equivalently, $\alpha_{IR,2\ell} = -\bar b_1/\bar b_2 = - 4\pi b_1/b_2$.  For
small enough fermion content, $b_2$ is positive, but as one enlarges the
fermion content in the theory, still retaining the property of asymptotic
freedom, the sign of $b_2$ can become negative, thereby producing an infrared
zero in $\beta_{\alpha,2\ell}$ at the above value.  If this occurs, then, as
the reference scale $\mu$ decreases from large values in the ultraviolet,
$\alpha(\mu)$ increases toward this infrared zero.  If the IR zero occurs at
sufficiently weak coupling, then one expects that the theory evolves from the
UV to the IR without confinement or spontaneous chiral symmetry breaking
(S$\chi$SB), to a non-Abelian Coulomb phase.  In this case, the infrared zero
of the beta function is an exact IR fixed point (IRFP) of the renormalization
group.  This was discussed for vectorial gauge theories in \cite{b2,bz}.

In a chiral gauge theory whose UV to IR evolution leads to a gauge coupling
that becomes strong in the infrared, there are several tools that one may use
in studying the possible resultant nonperturbative behavior.  First, one may
investigate whether the fermion content of the theory satisfies the 't Hooft
anomaly-matching conditions.  To do this, one determines the global flavor
symmetry group under which the theory is invariant and then examines various
candidate operator products for gauge-singlet composite spin 1/2 fermions to
ascertain if these can match the anomalies in the global flavor symmetries. If
this necessary condition is satisfied, then one possibility is that in the
infrared the strong chiral gauge interaction may confine and produce massless
composite spin 1/2 fermions (as well as massive gauge-singlet hadrons).  

An alternative possibility in a strongly coupled chiral gauge theory is that
the gauge interaction can produce bilinear fermion condensates.  In an
irreducibly chiral theory (without a vectorlike subsector), these condensates
break the gauge symmetry, as well as global flavor symmetries.  A method that
has been widely used to predict which type of condensate is most likely to form
in this case is the most-attractive-channel approach \cite{mac}.  Thus,
consider a bilinear fermion condensation channel in which fermions in the
representations $R_1$ and $R_2$ of a given gauge group form a condensate that
transforms according to the representation $R_{cond.}$ of this group, denoted
$R_1 \times R_2 \to R_{cond.}$.  An approximate measure, based on one-gluon
exchange, of the attractiveness of this condensation channel, is
\beq
\Delta C_2 \equiv C_2(R_1) + C_2(R_2) -C_2(R_{Ch}) \ ,
\label{deltac2}
\eeq
where $C_2(R)$ is the quadratic Casimir invariant for the representation $R$
(see Appendix \ref{group_invariants}) and $R_{Ch} \equiv R_{cond.}$.  At this
level of one-gluon exchange, if $\Delta C_2$ is positive (negative), then the
channel is attractive (repulsive). The most attractive channel is the one with
the largest (positive) value of $\Delta C_2$. According to the MAC approach, if
several possible condensation channels might, {\it a priori}, occur, then the
one that actually occurs is the channel that has the largest (positive) value
of $\Delta C_2$.  The MAC approach is supported by the fact that in quantum
chromodynamics, of the four {\it a priori} possible (Lorenz-invariant) bilinear
quark condensation channels $3 \times \bar 3 \to 1 + 8$ and 
$3 \times 3 \to \bar 3_a + 6_s$, 
only one occurs, and this is precisely the MAC, namely $3
\times \bar 3 \to 1$. The MAC method has been used in theoretical studies of
abstract chiral gauge theories and in efforts to build reasonably UV-complete
models with dynamical electroweak symmetry breaking \cite{uvcomplete_etc}.

An analysis of the Schwinger-Dyson equation for the propagator of a massless
fermion transforming according to the representation $R$ of a gauge group $G$
shows that, in the ladder (i.e., iterated one-gluon exchange) approximation the
minimum value of $\alpha$ for which fermion condensation occurs in a vectorial
gauge theory is given by the condition that $3\alpha_{cr} C_2(R)/\pi \sim 1$,
or equivalently, $3\alpha_{cr} \Delta C_2 /(2\pi) = 1$, since 
$\Delta C_2=2C_2(R)$ in this case \cite{chipt}.  Therefore, a rough estimate
for the minimal value of the running coupling which is sufficient to cause
condensation in a given channel $Ch$ is 
\beq
\alpha_{cr,Ch} \sim \frac{2\pi}{3 \Delta C_2(R)_{Ch}} \ ,
\label{alfcrit}
\eeq
Because of the strong-coupling nature of the fermion condensation
process, Eq. (\ref{alfcrit}) is only a rough estimate.  A measure of the
likelihood that the coupling grows large enough in the infrared to produce
fermion condensation in a given channel $Ch$ is thus the ratio
\beq
\rho_{_{IR,Ch}} \equiv \frac{\alpha_{IR,2\ell}}{\alpha_{cr,Ch}} \ .
\label{rho}
\eeq
If this ratio is significantly larger (smaller) than unity, one may infer that
condensation in the channel $Ch$ is likely (unlikely). As with the use of the
MAC, this $\rho$ test is only a rough estimate.  

In the case of evolution from the UV to an IR non-Abelian Coulomb phase, the
perturbative field degrees of freedom remain the same.  In the other types of
UV to IR evolution, in general, they change.  Given that one restricts to
asymptotically free theories, it is always possible to enumerate these field
degrees of freedom in the ultraviolet, and in many theories, one can also
enumerate them in the infrared.  Denoting the UV and IR measures as $f_{UV}$
and $f_{IR}$, it was conjectured that $f_{UV} \ge f_{IR}$ for vectorial gauge
theories in \cite{dfvgt}, and this conjecture was extended to chiral gauge
theories in \cite{dfcgt}, and further studied in \cite{ads,cgt} and in our
previous work, \cite{cgt2}, where several classes of chiral gauge theories were
constructued and shown to yield further support for this conjectured
inequality.  

Large-$N$ methods have also proved fruitful for the analysis of chiral gauge
theories that form families extendable to infinite $N$ \cite{eppz}.  However,
as we will show, the families of $S_k \bar A_\ell$ chiral gauge theories that
we construct and study here are only asymptotically free for a finite set of
$N$ values.  Going beyond the various approaches discussed in this section, one
would ideally hope to make use of fully nonperturbative methods that can be
used for any $N$, such as a lattice formulation and numerical simulations.
However, while the lattice formulation has been of great value for vectorial
gauge theories such as quantum chromodynamics, it has been difficult to use
lattice methods to study chiral gauge theories, owing to the presence of
fermion doubling.  


\section{Anomaly Freedom of a Low-Energy Effective Theory Arising from
  Dynamical Breaking of a Chiral Gauge Theory }
\label{descendant_anomalyfree}

Here we prove a general theorem that guarantees that a low-energy effective
field theory that arises from an anomaly-free chiral gauge theory via dynamical
gauge symmetry breaking is also anomaly-free.  Let us consider a chiral gauge
theory with a gauge group $G$ and an anomaly-free set of chiral fermions
transforming according to some set of representations $\{ R_i \}$ of $G$.
Without loss of generality, we take all of the fermions to be left-handed.
Also without loss of generality, we assume that this theory is irreducibly
chiral, i.e., does not contain any vectorlike subsector.  This assumption does
not entail any loss of generality because the fermions in a vectorlike
subsector give zero contribution to a chiral anomaly. Because the theory is
irreducibly chiral, the gauge symmetry precludes any fermion mass terms in the
fundamental lagrangian.  To begin with, we assume that $G$ is a simple group
and discuss later the straightforward generalization of our argument to the
case where $G$ is a direct-product group.  Let us denote the contribution of a
chiral fermion in the $R_i$ representation to the triangle anomaly in gauged
currents as ${\cal A}(R_i)$. The property that the initial theory is
anomaly-free is the condition
\beq
\sum_i n_{R_i} {\cal A}(R_i)=0 \ , 
\label{anomalycancellation_g}
\eeq
where $n_{R_i}$ denotes the number of copies of fermions in the representation
$R_i$.  This anomaly cancellation condition (\ref{anomalycancellation_g}) also
implies that if one restricts to a subgroup $H \subset G$, which means
decomposing each representation $R_i$ into representations $R_i'$ of $H$, then
the sum of contributions is also zero.  Now, assume that this theory is
asymptotically free, so that as the Euclidean reference scale $\mu$ decreases
from the UV to the IR, the running gauge coupling increases, and assume further
that this gauge coupling becomes strong enough at a scale $\Lambda$ to produce
bilinear fermion condensates that break the original gauge symmetry $G$ to a
subgroup $H \subset G$. The fermions involved in the condensate gain dynamical
masses of order $\Lambda$, and the gauge bosons in the coset $G/H$ also gain
masses of this order.

To construct the low-energy effective field theory that describes the physics
as the reference scale $\mu$ decreases below $\Lambda$, one integrates out
these massive states and enumerates the remaining $H$-nonsinglet massless
fields.  This enumeration involves decomposing each fermion representation
$R_i$ of $G$ in terms of representations $R_i'$ of $H$. The resultant anomaly
cancellation condition in the low-energy effective theory that is the
descendant of the original theory is
\beq
\sum_i n_{R_i'} {\cal A}(R_i')=0 \ , 
\label{anomalycancellation_h}
\eeq
where here ${\cal A}(R_i')$ refers to the contribution to the anomaly in the
descendant theory from the fermions in the $R_i'$ representation of the gauge
group $H$. Now the fermions in the original theory that were involved in the 
condensate, and hence acquired dynamical masses and were integrated out,
transform as singlets under $H$, and therefore, even if they were included in
Eq. (\ref{anomalycancellation_h}), they would make zero contribution to this
sum. Combining this fact with Eq. (\ref{anomalycancellation_g}), we deduce that
the remaining $H$-nonsinglet fermions must also make zero net contribution in
Eq. (\ref{anomalycancellation_h}).  This proves the theorem.  

We make some further remarks on this result.  In general, an asymptotically
free chiral gauge theory that becomes strongly coupled and produces fermion
condensates that dynamically break the gauge symmetry may undergo not just
one, but several sequential stages of dynamical gauge symmetry breaking.
Clearly, the theorem above applies not just to the first stage, but also to
subsequent stages of symmetry breaking. As noted, it is straightforward to
extend this theorem to the case where the gauge group of the theory is
a direct-product group instead of a simple group. An example of this is given
below in our analysis of the low-energy effective ${\rm SU}(5) \otimes {\rm
  U}(1)$ theory resulting as a descendant from an initial (anomaly-free) SU(6)
chiral gauge with fermions in the $S_2$ and $\bar A_2$ representations of
SU(6).

We next contrast our theorem with the situation concerning chiral gauge
symmetry breaking and associated fermion mass generation by the vacuum
expectation value (VEV) of a Higgs field.  For example, consider the Standard
Model, with gauge group $G_{SM}={\rm SU}(3) \otimes {\rm SU}(2) \otimes
{\rm U}(1)_Y = {\rm SU}(3) \otimes G_{EW}$ and Higgs field $\phi$.  In the
fundamental Lagrangian, the chiral $G_{EW}$ gauge symmetry forbids any mass
terms for the SM quarks and (SM-nonsinglet) leptons \cite{nur}.  Arranging the
Higgs potential $V(\phi)$ to have a minimum with a nonzero vacuum
expectation value of the Higgs field, $\langle \phi \rangle_0 = {0 \choose
  v/\sqrt{2}}$, breaks $G_{EW}$ to electromagnetic U(1)$_{em}$.  The SM quarks
and leptons gain masses through Yukawa interactions with the SM Higgs boson
$\phi$, via its nonzero vacuum expectation value.  For example, consider the
quarks.  Denote the left-handed quark fields as
\beq
Q_{i,L} \equiv { u_i \choose d_i}_L \ , \quad i=1,2,3 \ , 
\label{qil}
\eeq
and the right-handed quarks as $u_{i,R}$ and $d_{i,R}$, 
where here $i$ is a generational index, and we suppress color indices.  The
Yukawa term in the SM that generates the mass matrix for the up-type 
$q=2/3$ quarks is
\beq
{\cal L}_{Y,u}=- \sum_{i,j=1}^3 \bar Q_{i,L} Y^{(u)}_{ij}u_{j,R} \tilde \phi + 
h.c. \ , 
\label{yu}
\eeq
where $\tilde \phi \equiv i \sigma_1 \phi^*$
This yields the mass matrix for the charge $q=2/3$ quarks,
\beq
M^{(u)}_{ij} = \frac{v}{\sqrt{2}} \, Y^{(u)}_{ij} \ . 
\label{md}
\eeq
The diagonalization of this matrix yields the mass eigenstates for these
quarks.  For simplicity, assume that the Yukawa matrix is diagonal:
$Y^{(u)}_{ij}=Y^{(d)}_{ii}\delta_{ij}$. One could, formally at least, 
envision taking one element of this matrix to be arbitrarily large, say
$Y^{(u)}_{33} \gg 1$, so that the $t$ quark would have a mass $m_t
=M^{(u)}_{33} \gg v$.  If one were to attempt to integrate out the top quark,
the resulting theory would, at a perturbative level, appear to have gauge
anomalies (as well as an incomplete third quark generation).  The theorem that
we have proved shows that this sort of complication never happens in a chiral
gauge theory (without scalar fields) in which the gauge symmetry breaking and
fermion mass generation is dynamical, due to the formation of bilinear fermion
condensates. In the Higgs-Yukawa framework, if one were formally to take a
Yukawa coupling to infinity, the problem of the apparently anomalous low-energy
effective theory would be circumvented by the appearance of a nonperturbative
topological Wess-Zumino term in the action \cite{wz}.  However, there are
complications with trying to take a Yukawa coupling to be arbitrarily large,
because the beta function for the Yukawa coupling has an infrared zero at the
value zero (the ``triviality'' property of Yukawa theories), as was shown by
nonperturbative lattice measurements \cite{yr} and perturbative beta
function calculations (e.g., \cite{srgt} and references therein). 


\section{$S \bar A$ Theories}
\label{sabarmodel}

\subsection{Basic Construction} 

In this section we construct and analyze an interesting set of asymptotically
free chiral gauge theories with an SU($N$) gauge group and chiral fermions
transforming according to the rank-2 symmetric and conjugate antisymmetric
tensor representations of this group.  We denote these fermions generically as
$S_2$, and $\bar A_2$ (suppressing possible flavor indices) and equivalently by
the corresponding Young tableaux, $S_2 = \sym$ and $\bar A_2 =
\overline{\asym}$. To keep the notation as
simple as possible, we omit the subscripts where no confusion will result, 
setting
\beq
S_2 \equiv S, \quad \bar A_2 \equiv \bar A \ .
\label{s2sa2a}
\eeq
We denote the number of $S$ and $\bar A$ fields as $n_S$ and $n_{\bar A}$. An
$S \bar A$ theory is irreducibly chiral, i.e., it does not contain any
vectorial subset.  The chiral gauge symmetry forbids any fermion mass term in
the Lagrangian.  The triangle anomaly $\cal A$ in gauged currents of our $S
\bar A$ theory is
\beqs
{\cal A} & = & n_S {\cal A}(S) + n_{\bar A}{\cal A}(\bar A) \cr\cr
         & = & n_S {\cal A}(S) - n_{\bar A}{\cal A}(A) \ . 
\label{anomsabar}
\eeqs
Substituting ${\cal A}(S) = N+4$ and ${\cal A}(A) = N-4$ (see Appendix
\ref{group_invariants}), the condition that this $S \bar A$ theory should be
free of a triangle anomaly in gauged currents is that
\beq
n_S(N+4)- n_{\bar A}(N-4) = 0 \ . 
\label{anomalycancellation}
\eeq
Thus, $n_S$ and $n_{\bar A}$ take values in the ranges $n_S \ge 1$ and 
$n_{\bar A} \ge 1$, subject to the anomaly cancellation condition
(\ref{anomalycancellation}) and the requirement that the resultant $S \bar A$
theory must be asymptotically free.  A member of this set of chiral gauge
theories is thus defined as
\beq
S \bar A: \quad G={\rm SU}(N), \quad {\rm fermions}: \ n_S S + 
n_{\bar A} \bar A \ , 
\label{sabmodel}
\eeq
where it is understood implicitly that $N$, $n_S$, and $n_{\bar A}$ satisfy
the condition (\ref{anomalycancellation}) and yield an asymptotically free
theory. We denote such a theory, for short, as $(N;n_S,n_{\bar A})$.

The anomaly cancellation condition (\ref{anomalycancellation}) is a linear
diophantine equation.  If $N=4$, i.e., $G={\rm SU}(4)$, a nontrivial solution
of this equation is not possible, because in this case the $\asym$
representation is self-conjugate, and hence has zero anomaly, so there is no
value of $n_{\bar A}$ which can cancel the contribution to the anomaly in
gauged currents from the fermions in the $\sym$ representation.  Consequently,
a nontrivial solution of the anomaly cancellation condition
(\ref{anomalycancellation}) requires that $N \ge 5$, and we restrict to this
range.  We define the ratio
\beq
\frac{n_{\bar A}}{n_S} = \frac{N+4}{N-4} \equiv p \ . 
\label{p}
\eeq
Since the right-hand side of Eq. (\ref{p}) is greater
than one, it follows that $n_S < n_{\bar A}$. Therefore, the theories of this
type with minimal chiral fermion content have $n_S=1$ and take the form
\beq
(N;n_S,n_{\bar A}) = (N;1,p) \ , 
\label{nabar}
\eeq
with the understanding that $n_{\bar A}$ must be a (positive) integer. We find
that there are precisely four solutions of Eq. (\ref{p}) with $n_S=1$ that
satisfy this condition, namely (including the $N$ characterizing the SU($N$)
gauge group)
\beqs
& & (N;n_S,n_{\bar A}) = (5;1,9), \ (6;1,5), \ (8;1,3), \ (12;1,2) \ . \cr\cr
& & 
\label{minimalfermions}
\eeqs

In the context of anomaly cancellation alone, before imposing the condition of
asymptotic freedom, we observe a basic mathematical property. If
$(N;n_S,n_{\bar A})$ is a solution of Eq. (\ref{anomalycancellation}), then a
theory with $n_{cp}$ copies (abbreviated $cp$) of the fermion content also
yields a solution of (\ref{anomalycancellation}).  That is,
\beqs
& & (N;n_S,n_{\bar A}) \ {\rm is \ anom. \ free} \ \Longrightarrow \cr\cr
& & (N;n_{cp}n_S,n_{cp}n_{\bar A}) 
\ {\rm is \ anom. \ free \ for } \ n_{cp} \ge 2.
\label{ncp}
\eeqs
If one were not to require that the theory must be asymptotically free, then
$n_{cp}$ could be any positive integer, and hence the linear diophantine
equation (\ref{anomalycancellation}) would have an infinite number of
solutions. However, we do require that our chiral gauge theories must be
asymptotically free so that they are perturbatively calculable in the deep 
ultraviolet.  

Given this requirement, the next step is to ascertain, for a
given value of $N$, which values of $n_{cp}$ are allowed by asymptotic freedom.
To do this, we calculate the first two coefficients of the beta function. These
coefficients are
\beq
b_1 = \frac{1}{3}\bigg [11N- \Big \{ n_S(N+2)+n_{\bar A}(N-2) \Big \} \bigg ]
\label{b1nsnab}
\eeq
and 
\beqs
b_2 & = & \frac{1}{3}\bigg [ 34N^2 
- n_S       \bigg \{ 5N+3\frac{(N+2)(N-1)}{N} \bigg \}(N+2) \cr\cr
& & 
- n_{\bar A}\bigg \{ 5N+3\frac{(N-2)(N+1)}{N} \bigg \}(N-2) \ \bigg ] 
\ . 
\label{b2nsnab}
\eeqs
It will be useful to reexpress these coefficients in a convenient form for
analysis of the minimal set of fermions, viz., $(n_S,n_{\bar A})=(1,p)$, given
explicitly in (\ref{minimalfermions}) and the sets involving $n_{cp}$-fold
replication (copies, or flavors) of minimal sets,
\beq
(n_S,n_{\bar A})=n_{cp}(1,p) = (n_{cp},n_{cp}p) \ . 
\label{ncpsols}
\eeq
Thus, equivalently, 
\beqs
b_1 &=&\frac{1}{3}\bigg [11N-n_{cp}\Big \{ (N+2)+p (N-2) \Big \} \ \bigg ] 
\cr\cr
    &=& \frac{1}{3}\bigg [ 11N - 2n_{cp}\Big ( \frac{N^2-8}{N-4} \Big ) 
\ \bigg ]
\label{b1ncopy}
\eeqs
and
\newpage
\beqs
b_2 & = & \frac{1}{3}\bigg [ 34N^2
-n_{cp}\bigg \{ \Big ( 5N+3\frac{(N+2)(N-1)}{N} \Big )(N+2) \cr\cr
& &          + p\Big ( 5N+3\frac{(N-2)(N+1)}{N} \Big )(N-2) \bigg \} \ \bigg ]
 \cr\cr
& = & \frac{1}{3}\bigg [ 34N^2 -8n_{cp}\Big \{ \frac{2N^4-19N^2+12}
{N(N-4)} \Big \} \ \bigg ] \ , \cr\cr
& &
\label{b2ncopy}
\eeqs
where here it is understood that, since $n_S$ is taken to have its minimal
value of 1, the value of $N$ is restricted so that $p$ is
a (positive) integer.  The requirement of asymptotic freedom, i.e., $b_1 > 0$,
implies that $n_{cp}$ is bounded above according to
\beq
n_{cp} < \frac{11N(N-4)}{2(N^2-8)} \ .
\label{ncopy_upperbound}
\eeq

As a rational number, this upper bound has the respective values (quoted to the
indicated floating-point accuracy) 1.62, \ 2.36, \ 3.14, and 3.88 for $N=5$, 
\ 6, \ 8, \ 12.  Therefore, on the integers, we have the upper bounds 
\beq
n_{cp} \le \cases{ 1 & for $N=5$ \cr
                   2 & for $N=6$ \cr
                   3 & for $N=8, \ 12$ }
\label{ncopymax}
\eeq
Thus, the full set of (anomaly-free) asymptotically free $S \bar A$ chiral 
gauge theories of this type, $(N;n_S,n_{\bar A})=(N;1,p)$ and 
$(N;n_{cp},n_{cp}p)$ with integer $p$, includes, in addition to the minimal
set (\ref{minimalfermions}), also the additional theories with an $n_{cp}$-fold
replication of the set (\ref{minimalfermions}), namely
\beqs
(N;n_S,n_{\bar A}) & = & (6;2,10), \ (8;2,6), \ (8;3,9), \cr\cr
                   & & (12;2,4), \ (12;3,6) \ . 
\label{ncopyfermions}
\eeqs

There are also (asymptotically free) solutions of the anomaly cancellation
condition (\ref{anomalycancellation}) with nonminimal values $n_S > 1$ that are
not of the form of simple replications of the minimal set
(\ref{minimalfermions}), i.e., for which $p$ is a (positive) rational, but not
integer, number.  We find that there are seven such solutions, namely 
\beqs
(N;n_S,n_{\bar A}) & = & (10;3,7), \ (16;3,5), \ (20;2,3), \ (20;4,6), \cr\cr
& & (28;3,4), \ (36;4,5), \ (44;5,6) \ . 
\label{othersols}
\eeqs
Thus, we find that there are sixteen $S \bar A$ anomaly-free asymptotically
free chiral gauge theories; these consist of the four minimal ones of the form
$(N;1,p)$ in Eq. (\ref{minimalfermions}), the five theories of the
form $(N;n_{cp},n_{cp}p)$ in 
Eq. (\ref{ncopyfermions}), and the seven additional ones in
Eq. (\ref{othersols}) with rational, but non-integral $p$.  As noted in the
introduction, a striking feature
of this family of $S \bar A$ chiral gauge theories is that the combined
requirements of anomaly cancellation and asymptotic freedom yields only a
finite set of solutions, in contrast to generic families of chiral gauge
theories that have been studied in the past, such as $S + (N+4)\bar F$
and $A + (N-4)\bar F$, and extensions of these with vectorlike subsectors, 
which allow, respectively, the infinite ranges $N \ge 3$ and $N \ge 5$. 

We label the fermion fields in a given $S \bar A$ theory as 
\beq
S_i: \ \psi^{ab}_{i,L} = \psi^{ba}_{i,L} \ , \quad 1 \le i \le n_S
\label{psi}
\eeq
and
\beq
\bar A_j: \ \chi_{ab,j,L} = - \chi_{ba,j,L}, 
\quad 1 \le j \le n_{\bar A} \ , 
\label{chi}
\eeq
where $a, \ b$ are SU($N$) gauge indices and $i, \ j$ are flavor indices. 


\subsection{Global Flavor Symmetry}

The classical global flavor ($fl$) symmetry group of the $(N;n_S,n_{\bar A})$
\ $S \bar A$ theory is
\beqs
& & G_{fl,cl} = {\rm U}(n_s) \otimes {\rm U}(n_{\bar A}) \cr\cr
         & = & \cases{ {\rm SU}(n_{\bar A}) \otimes {\rm U}(1)_S \otimes 
{\rm U}(1)_{\bar A} & if $n_s=1$ \cr
{\rm SU}(n_S) \otimes {\rm SU}(n_{\bar A}) \otimes {\rm U}(1)_S \otimes 
{\rm U}(1)_{\bar A} & if $n_s \ge 2$} \cr\cr
& & 
\label{gfcl}
\eeqs
The operation of the elements of these global groups on the fermion fields is
as follows.  For fixed SU($N$) group indices $a, \ b$, the theory is invariant
under the action of an element ${\cal U}_S \in {\rm U}(n_S)$ on the
$n_S$-dimensional vector $(\psi^{ab}_{1,L},
\psi^{ab}_{2,L},...\psi^{ab}_{n_S,L})$
\beq
\psi^{ab}_{i,L} \to \sum_{j=1}^{n_S} ({\cal U}_S)_{ij} \psi^{ab}_{j,L}
\label{sglobaltransformation}
\eeq
and separately under the action of an element of an element 
${\cal U}_{\bar A} \in {\rm U}(n_{\bar A})$ on the
$n_{\bar A}$-dimensional vector $(\chi_{ab,1,L},
\chi_{ab,2,L},...\chi_{ab,n_{\bar A},L})$
\beq
\chi_{ab,i,L} \to \sum_{j=1}^{n_{\bar A}} ({\cal U}_{\bar A})_{ij} 
\chi_{ab,j,L} \ . 
\label{abarglobaltransformation}
\eeq
The U(1)$_S$ and U(1)$_{\bar A}$ global symmetries are both broken by SU($N$)
instantons \cite{instantons}. As before in our analysis of different chiral
gauge theories \cite{cgt}, we define a vector whose components are comprised of
the instanton-generated contributions to the breaking of these symmetries. 
In the basis $(S,\bar A)$, this vector is
\beqs
{\vec v} & = & \Big ( n_S T(S), \ n_{\bar A} T(\bar A) \Big ) = 
 n_{np} \bigg ( \frac{N+2}{2}, \ p \Big ( \frac{N-2}{2} \Big ) \bigg ) \cr\cr
 & = & \frac{n_{np}}{2}\bigg (N+2, \ \frac{(N+4)(N-2)}{N-4} \bigg ) \ .  
\label{anomvec}
\eeqs
We can construct one linear combination of the two original currents that is
conserved in the presence of SU($N$) instantons. We denote the corresponding
global U(1) flavor symmetry as U(1)$^\prime$ and the fermion charges under this
U(1)$^\prime$ as
\beq
{\vec Q}' = \Big ( Q'_S, \ Q'_{\bar A} \Big ) \ . 
\label{qvec}
\eeq
The U(1)$^\prime$ current is conserved if and only if 
\beq
\sum_f n_f T(R_f) \, Q_f' = {\vec v} \cdot {\vec Q}' = 0 \ . 
\label{u1inv}
\eeq
Clearly, this condition determine the vector ${\vec Q}'$ only up 
to an overall multiplicative constant.  A solution is 
\beq
{\vec Q}' = \Big ( (N-2)(N+4), \ -(N+2)(N-4) \ \Big ) \ .
\label{qvecsol}
\eeq
The actual global chiral flavor symmetry group (preserved in the presence of
instantons) is then 
\beq
G_{fl} = \cases{ {\rm SU}(n_{\bar A}) \otimes {\rm U}(1)' & if $n_S=1$ \cr
 {\rm SU}(n_S) \otimes {\rm SU}(n_{\bar A}) \otimes {\rm U}(1)' & if 
$n_S \ge 2$} 
\label{gfl}
\eeq

So far, our discussion of global flavor symmetries applies generally to all of
the $S \bar A$ chiral gauge theories.  We next determine the most attractive
channel for fermion condensation, which differs for $N=5$ and $N \ge 6$, and
then proceed with analyses of specific $S \bar A$ theories. 


\subsection{Fermion Condensation Channels}

For $N \ne 5$, the most attractive channel for the formation of a bilinear
fermion condensate in a $(N;n_S,n_{\bar A})$ $S \bar A$ chiral gauge theory is 
\beq 
S \times \bar A \to adj \ , 
\label{sabtoadj}
\eeq
where $adj$ denotes the adjoint representation of SU($N$).  This has 
\beq
\Delta C_2 = \frac{(N+2)(N-2)}{N}  \quad {\rm for} \ 
S \times \bar A \to adj \ . 
\label{DeltaC2_sabtoadj}
\eeq
Substituting this expression for $\Delta C_2$ into Eq. (\ref{alfcrit}) for 
the estimate of the minimum critical coupling for condensation in this 
channel, we obtain
\beq
\alpha_{cr} \simeq \frac{2\pi N}{3(N+2)(N-2)} \quad {\rm for} \ 
S \times \bar A \to adj \ . 
\label{alfcrit_sabtoadj}
\eeq
In general, there are several stages of fermion condensation, as will be
evident in the analyses of specific theories below. 
In the $S \bar A$ theory with $G={\rm SU}(5)$, the most attractive channel is
$\bar A \times \bar A \to F$ instead of (\ref{sabtoadj}) and will be discussed
in the section devoted to this theory.


\section{ $S \bar A$ Theory with $G={\rm SU}(5)$ }
\label{su5}

\subsection{General} 

Our SU(5) $S \bar A$ theory has $(n_S,n_{\bar A})=(1,9)$.  Since $n_S=1$ for
this theory, we use a simplified notation without the flavor index on the $S$
field, namely $\psi^{ab}_{i=1,L} \equiv \psi^{ab}_L$. We recall that the
$S=\sym$ and $\bar A = \overline{\asym}$ representations of SU(5) have
dimensionalities 15 and 10, respectively, and we shall equivalently refer to
them in this section by these dimensionalities.  From Eq. (\ref{gfl}), this
theory has a (nonanomalous) global flavor symmetry
\beq
G_{fl} = {\rm SU}(9)_{\bar A} \otimes {\rm U}(1)' \ .
\label{gfl_Nc5ncopy1}
\eeq
We have not found SU(5)-gauge-singlet composite-fermion operators that satisfy
the 't Hooft anomaly matching conditions for this theory. Indeed, the minimal
fermionic operator products, such as $S^{ab}\bar A_{bc} S^{cd}$,
$\epsilon^{abcde} \bar A_{ab} \bar A_{cd} \bar A_{ef}$, are not SU(5) gauge
singlets. 

We list the values of the first two coefficients of the beta
function for this theory in Table \ref{minimal_andcopies}. This beta function
has an IR zero which occurs, at the two-loop level, at a value
$\alpha_{IR,2\ell}=0.645$.  As will be discussed further below, we find that
this value is close to an estimate of the minimal critical value, $\alpha_{cr}$
for the formation of a bilinear fermion condensate in the most attractive
channel, with associated spontaneous chiral symmetry breaking.  Consequently,
we shall analyze both possibilities (i.e., retaining or breaking chiral
symmetry) for the UV to IR evolution of this SU(5) $S \bar A$ theory.

The most attractive channel for condensation is
\beq 
{\rm MAC \ for} \ (5;1,9): \quad \overline{\asym} \times \overline{\asym}
\to \fund, \ i.e., \ \bar{A} \times \bar{A} \to F \ ,
\label{10bar10barto5}
\eeq
where $F=\fund$ is the fundamental representation.  Equivalently, in terms of
dimensionalities, this is the channel $\overline{10} \times \overline{10} \to
5$.  Since $C_2(A_2)=18/5$ and $C_2(F)=12/5$ for SU(5), the measure of
attractiveness for this channel is
\beq
\Delta C_2 = \frac{24}{5} \quad {\rm for} \ 
\overline{10} \times \overline{10} \to 5 \ . 
\label{DeltaC2_10bar10barto5}
\eeq
(The next-most attractive channel
is $S \times \bar A \to {\rm adj}$, with $\Delta C_2=21/5$.) The rough 
Schwinger-Dyson estimate of the critical coupling for condensate formation in
the $\overline{10} \times \overline{10} \to 5$ channel is 
$\alpha_{cr} \sim 5\pi/36 = 0.44$
To compare $\alpha_{IR,2\ell}$ with $\alpha_{cr}$, we use the ratio
$\rho$ defined in Eq. (\ref{rho}). We calculate $\rho=1.5$ for the channel
(\ref{10bar10barto5}).  This value of $\rho$ is close enough to unity that we
cannot make a definite conclusion concerning the presence or absence of
spontaneous chiral symmetry breaking. There are thus two possibilities for the
first step in the UV to IR evolution of this SU(5) $S \bar A$ theory, and we 
investigate both of these. 


\subsection{ Evolution of SU(5) $S \bar A$ Theory to a Non-Abelian Coulomb
  Phase in the IR}

First, the SU(5) (5;1,9) $S \bar A$ theory might evolve downward in $\mu$
without any spontaneous chiral symmetry breaking, yielding a (deconfined)
non-Abelian Coulomb phase (NACP) in the infrared. We denote this possibility as
\beq
(5;1,9): \quad {\rm UV} \to {\rm IR \ \ NACP} \ . 
\label{nacp}
\eeq
In this case, the global flavor symmetry in the IR is the same as in the UV,
namely (\ref{gfl_Nc5ncopy1}). 


\subsection{Dynamical Breaking of SU(5) to SU(4) Gauge Symmetry}

Second, the gauge coupling of the (5;1,9) SU(5) $S \bar A$ theory might become
sufficiently strong to lead to nonperturbative behavior.  Since we have not
found gauge-singlet operator products that satisfy 't Hooft anomaly matching
conditions in this SU(5) theory, we infer that this nonperturbative
behavior would lead to the formation of a bilinear fermion condensate, breaking
the SU(5) gauge symmetry. We denote this possibility as
\beq
(5;1,9): \quad {\rm UV} \to {\rm IR}: \quad S \chi SB \ \Longrightarrow 
{\rm SU}(4) \ . 
\label{su5tosu4}
\eeq
We proceed to analyze this possibility. Thus, we assume that as the reference
Euclidean momentum scale $\mu$ decreases below a value that we denote
$\Lambda_5$, the gauge coupling becomes large enough to form a bilinear
condensate in the most attractive channel, $\bar A \times \bar A \to F$,
Eq. (\ref{10bar10barto5}),
dynamically breaking the SU(5) gauge symmetry to SU(4) (and also breaking the
global flavor symmetry (\ref{gfl_Nc5ncopy1}) ).  The associated fermion
condensate is of the form 
$\langle \epsilon^{abdef}\chi_{bd,i,L}^T C \chi_{ef,j,L} \rangle$, where $C$ 
is the charge-conjugation Dirac matrix.  With no loss of generality, we may
choose the uncontracted index $a$ to be $a=5$.  By a vacuum alignment argument
similar to that used in \cite{cgt2}, we infer that the actual condensates are
of the form
\beqs
& & \langle \epsilon^{5bdef}\chi_{bd,j,L}^T C \chi_{ef,j,L} \rangle \propto 
\Big [ \langle \chi_{12,j,L}^T C \chi_{34,j,L} \rangle \cr\cr
& - & \langle \chi_{13,j,L}^T C \chi_{24,j,L} \rangle +  
\langle \chi_{14,j,L}^T C \chi_{23,j,L} \rangle \Big ] 
\cr\cr
& & 
\label{10bar10barto5condensate}
\eeqs
for $1 \le j \le 9$. Since the gauge interaction is independent of the flavor
index, it follows that these condensates have a common value independent of the
flavor index, $j$.  The fermions involved in these condensates thus gain a
common dynamical mass of order $\Lambda_5$ and are integrated out of the
low-energy effective field theory applicable at energy scales $\mu<\Lambda_5$.
When SU($N$) breaks to SU($N-1$) there are $2N-1$ gauge bosons
in the coset ${\rm SU}(N)/{\rm SU}(N-1)$, corresponding to the broken
generators.  Here, with $N=5$, there are nine gauge bosons in the coset
SU(5)/SU(4), and these also gain masses of order $\Lambda_5$.


\subsection{Analysis of SU(4) Descendant Theory }

Since the low-energy effective field theory resulting as a descendant from the
breaking of the SU(5) $S \bar A$ gauge symmetry is invariant under an SU(4)
gauge symmetry, in order to analyze it, we decompose the remaining massless
fermions into SU(4) representations.  For this purpose, we make use of the
following general results for an SU($N$) group:
\beq
\sym_{\ {\rm SU}(N)}=[ \ \sym + \fund + 1 \, ]_{\ {\rm SU}(N-1)} \ , 
\label{symdecomp}
\eeq
where 1 denotes a singlet, and, for $N \ge 4$, 
and
\beq
\overline{\asym}_{\ {\rm SU}(N)} = 
[ \ \overline{\asym} + \overline{\fund} \ ]_{\ {\rm SU}(N-1)} \ . 
\label{asymdecomp}
\eeq
Here, in terms of dimensionalities, these decompositions read 
\beq
15_{{\rm SU}(5)}= 10_{{\rm SU}(4)}+4_{{\rm SU}(4)}+1
\label{sym_su5tosu4decomp}
\eeq
and
\beq
\overline{10}_{{\rm SU}(5)}=6_{{\rm SU}(4)}+\bar{4}_{{\rm SU}(4)} \ .
\label{asym_su5tosu4}
\eeq
Note that $6_{{\rm SU}(4)}$ is self-conjugate, i.e., 
$6_{{\rm SU}(4)} \approx \bar{6}_{{\rm SU}(4)}$. 
The massless SU(4)-nonsinglet fermions in the SU(4) theory thus consist
of $\psi^{ab}_L$ with $1 \le a, \ b \le 4$, $\psi^{a5}_L$ with $1 \le a \le 4$,
and $\chi_{a5,j,L}$ with $1 \le a \le 4$ and $1 \le j \le 9$.  In terms of
Young tableaux, these are $\sym + \fund + 9 \, \overline{\fund}$ under SU(4), 
or in equivalent notation, the theory is 
\beq
{\rm SU}(4), \ {\rm fermions}: \ S + F + 9 \bar F = 
S + 8\bar F + 1\{F + \bar F \} 
\label{su4fromsu5}
\eeq
Thus the SU(4)-nonsinglet fermion content of the theory is precisely the $N=4$,
$p=1$ special case of the $Sp$ model presented in \cite{by} and further
analyzed in \cite{by2,ads,cgt}, so we can apply the results from these previous
studies here.  This SU(4) theory also contains the massless SU(4)-singlet
chiral fermion $\psi^{55}_L$ inherited from the SU(5) theory, but this does not
affect the SU(4) dynamics.  We recall that the $Sp$ model is defined by the
gauge group and chiral fermion content
\beq
Sp: \ G={\rm SU}(N), \ {\rm fermions}: \ S + (N+4)\bar F + p\{F + \bar F\}
\label{spmodel}
\eeq
where the first part, $S + (N+4)\bar F$ is irreducibly chiral and the second
part is a vectorlike subsector consisting of $p$ copies of $\{F + \bar F \}$.
As dictated by our theorem proved above, this SU(4) descendant theory is
anomaly-free. This is evident from the count
\beqs
{\cal A}(\sym_{\ {\rm SU}(4)}) & + & {\cal A}(\fund_{\ {\rm SU}(4)}) 
- 9 \, {\cal A}(\fund_{\ {\rm SU}(4)}) \cr\cr
& = & 8+1-9=0 \ . 
\label{su4anom}
\eeqs

The first two coefficients of the beta function of this SU(4) low-energy 
effective field theory have the same sign (explicitly, $\bar b_1=0.7427$
and $\bar b_2=0.1831$), so this beta function has no IR zero at the maximal
scheme-independent, two-loop order. Thus, as the Euclidean momentum scale $\mu$
decreases below $\Lambda_5$, the SU(4) gauge coupling inherited from the
original SU(5) theory continues to increase until it exceeds the region where
it can be described by the perturbative beta function.  There are then several
possibilities for the next stage of RG evolution to lower scales. We discuss
these next.


\subsubsection{ Confinement in SU(4) Theory with Massless Composite Fermions}

The first of these possibilities for the SU(4) theory is present because of the
fact that (for general values of $N$ and $p$ where there is confinement) the
$Sp$ model satisfies the 't Hooft anomaly-matching conditions
\cite{by,ads,cgt}.  Owing to this, as the gauge coupling continues
to increase in the infrared, the gauge interaction could confine the (massless)
SU(4)-nonsinglet fermions, producing massless spin 1/2 composite fermions as
well as massive SU(4)-singlet hadrons (mesons, glueballs, and mass eigenstates
that are linear combinations of mesons and glueballs).  The massless fermion
spectrum would also contain the SU(4)-singlet chiral fermion $\psi^{55}_L$ from
the original SU(5) $S \bar A$ theory.


\subsubsection{ Formation of Fermion Condensates Breaking SU(4) Gauge Symmetry}

The second of these possibilities for the UV to IR evolution of the SU(4)
low-energy effective field theory resulting from the breaking of the SU(5) 
$S \bar A$ theory is further fermion condensation in the most attractive 
channel in this SU(4) theory. The MAC is the channel 
$\sym \times \overline{\fund} \to \fund$, i.e.,
\beq
S \times \bar F \to F \ . 
\label{sfbarchannel}
\eeq
The next-most attractive channel is $F \times \bar F \to 1$, with 
\beq
\Delta C_2 = 2C_2(F) = \frac{N^2-1}{N} \ . 
\label{DeltaC2_FFbar}
\eeq
The fact that the $S \times \bar F \to F$ channel is the
MAC is evident from the property that it has a larger $\Delta C_2$ value than
the $F \times \bar F \to 1$ channel:
\beq
\frac{(N+2)(N-1)}{N} - \frac{N^2-1}{N} = \frac{N-1}{N} > 0 \ . 
\label{diffDeltaC2}
\eeq
For generality, we discuss the physics of the $S \times \bar F \to F$ channel
for general $N$, although our specific application will be to $N=4$. 
The attractiveness measure for this channel is 
\beq
\Delta C_2 = C_2(S) = \frac{(N+2)(N-1)}{N} \quad {\rm for} \ 
S \times \bar F \to F \ . 
\label{DeltaC2_SFbar}
\eeq
Substituting this into Eq. (\ref{alfcrit}) for the estimate of the minimum 
critical coupling for condensation in this channel, we obtain
\beq
\alpha_{cr} \simeq \frac{2\pi N}{3(N+2)(N-1)} \quad {\rm for} \ 
S \times \bar F \to F \ . 
\label{alfcrit_sfbar}
\eeq
For present case of $N=4$, this yields the estimate $\Delta C_2 = 4.5$.  We
denote the Euclidean scale $\mu$ at which the running coupling $\alpha(\mu)$
exceeds the critical value for condensation in this MAC as $\Lambda_4$.  The
condensation (\ref{sfbarchannel}) breaks SU(4) to SU(3). The associated
condensate has the general form $\langle \sum_{b=1}^4 \psi^{ab \ T}_L C
\chi_{5b,j,L}\rangle$.  Without loss of generality, we can denote the breaking
axis as $a=4$ and label the copy (flavor) index of the $\bar F$ fermion
$\chi_{5b,j,L}$ involved in this condensate as $j=9$, so that the condensate is
\beq
\langle \sum_{b=1}^4 \psi^{4b \ T}_L C \chi_{b5,9,L}\rangle \ . 
\label{su4condensate}
\eeq
The fermions $\psi^{b4}_L$ and $\chi_{b5,9,L}$ with $1 \le b \le 4$ involved in
this condensate thus get common dynamical masses of order $\Lambda_4$.  The
seven gauge bosons in the coset SU(4)/SU(3) also get masses of order
$\Lambda_4$.  These fermions and bosons are integrated out of the low-energy
effective field theory that is operative for scales $\mu < \Lambda_4$.

This low-energy effective field theory is invariant under an (anomaly-free)
SU(3) gauge symmetry and contains the massless SU(3)-nonsinglet chiral fermions
$\psi^{ab}_L$ with $1 \le a, \ b \le 3$, transforming as $S = \sym$ of SU(3),
$\psi^{a5}_L$, with $1 \le a \le 3$, transforming as $F = \fund$ of SU(3), and
the $\chi_{a5,j,L}$ with $1 \le a \le 3$ and $1 \le j \le 8$; that is,
\beq
G={\rm SU}(3), \ {\rm fermions}: \quad S + F + 8 \, \bar F 
= S + 7\bar F + 1\{F + \bar F \} 
\label{su3from su4}
\eeq
The SU(3)-nonsinglet fermion content of this theory is the $N=3$, $p=1$ special
case of the $Sp$ model (\ref{spmodel}).  This SU(3) theory also contains a
number of massless SU(3)-singlet chiral fermions. In addition to the
$\psi^{55}_L$ SU(4)-singlet fermion remaining from the ${\rm SU}(5) \to {\rm
  SU}(4)$ breaking at the higher scale $\Lambda_5$, there are also nine
massless SU(3)-singlet fermions remaining from the 
${\rm SU}(4) \to {\rm   SU}(3)$ breaking at $\Lambda_4$, namely 
$\psi^{45}_L$ and the $\chi_{45,j,L}$ with $1 \le j \le 8$.

As discussed in \cite{by,ads,cgt}, the further evolution into the infrared of
this SU(3) $Sp$ model might lead to confinement with resultant massless
composite fermions or to further condensation in the most
attractive channel, which is $S \times \bar F \to F$, breaking SU(3) to SU(2)
and then breaking SU(2) completely. In the latter case, the full sequence of
gauge symmetry breaking of (5;1,9) theory would be as follows: 
$\bar A \times \bar A \to F$, breaking SU(5) to SU(4), followed
in the resultant SU(4) descendant theory by the condensation 
$S \times \bar F \to F$, breaking SU(3) to SU(2), followed again by
condensation in the respective $S \times \bar F \to F$ channel, breaking 
SU(2) completely. 


\section{ $S \bar A$ models with $N \ge 6$ }
\label{sabar_nge6}

\subsection{General Analysis} 

We next proceed to analyze the $S \bar A$ models $(N;n_S,n_{\bar A})$ with $N
\ge 6$.  In contrast to the SU(5) $S \bar A$ theory $(5;1,9)$, if $N \ge 6$,
the most attractive channel for bilinear fermion condensation is 
$S \times \bar A \to adj$, as given in Eq. (\ref{sabtoadj}):
This condensation produces, as the first stage of dynamical gauge symmetry
breaking, the pattern 
\beq
{\rm SU}(N) \to {\rm SU}(N-1) \otimes {\rm U}(1) \ .
\label{sabar_nge6_symmetrybreaking}
\eeq
The values of $\Delta C_2$ and $\alpha_{cr}$ for this channel were given in
Eqs. (\ref{DeltaC2_sabtoadj}) and (\ref{alfcrit_sabtoadj}). The resultant
estimates for $\alpha_{cr}$ for condensation in this channel in specific
$(N;n_S,n_{\bar A})$ models with $N \ge 6$ are listed in Tables
\ref{minimal_andcopies} and \ref{other_theories}.  In these tables we also list
the (reduced) beta function coefficients $\bar b_1$ and $\bar b_2$, the
resultant IR zero in the two-loop beta function, if it exists, and the ratio
$\rho=\alpha_{IR,2\ell}/\alpha_{cr}$ from Eq. (\ref{rho}). In cases where the
beta function has no IR zero, the coupling increases with decreasing reference
scale $\mu$ until it exceeds the perturbatively calculable regime.  For a
generic $S \bar A$ $(N;n_S,n_{\bar A})$ theory, we do not find solutions for 't
Hooft anomaly matching, although, as will be discussed later, in certain cases,
resultant low-energy effective descendant field theories with different fermion
content (e.g., the SU(4) $Sp$ model below), do satisfy these matching
conditions.  Thus, as regards the initial UV to IR evolution of the
$(N;n_S,n_{\bar A})$ theory, if the gauge coupling becomes sufficiently strong,
one expects fermion condensation.  The resulting expectations for whether or
not fermion condensation and associated spontaneous chiral symmetry breaking
occur are listed for the sixteen $S \bar A$ theories in Tables
\ref{minimal_andcopies} and \ref{other_theories}.


\subsubsection{Flow to Chirally Symmetric Non-Abelian Coulomb Phase in IR} 

Referring to these Tables \ref{minimal_andcopies} and \ref{other_theories}, in
the six cases where the value of $\rho$ is substantially less than unity, we
infer that the theory is likely to evolve smoothly from the UV to a
(deconfined) chirally symmetric non-Abelian Coulomb phase in the
IR. Explicitly, we infer that this IR behavior occurs for the
(6;2,10), \ (8;3,9), \ (10;3,7), \ (20;4,6), \ (36;4,5), \ and (44;5,6) $S \bar
A$ theories. 


\subsubsection{Flow to IR with Spontaneous Chiral Symmetry Breaking}

We next discuss the situation in which, as the reference scale $\mu$ decreases
from the UV to the IR, the coupling becomes large enough so that
nonperturbative behavior occurs.  As noted, in the absence of sets of fermionic
operator products that yield solutions to 't Hooft anomaly matching conditions,
one infers that this nonperturbative behavior entails fermion condensation and
associated spontaneous breaking of the SU($N$) gauge symmetry (although after
some stage(s) of such symmetry breaking, a low-energy descendant theory may
satisfy these matching conditions).  For technical simplicity, we restrict our
discussion to the minimal theories $(N;1,p)$; corresponding analyses can be
given for the other $(N;n_S,n_{\bar A})$ models.  As is evident from Table
\ref{minimal_andcopies}, all three of of the $(N;1,p)$ theories with $N \ge 6$,
namely (6;1,5), (8;1,3), and (12;1,2) have the property that the gauge coupling
becomes sufficiently strong to produce further bilinear fermion condensation.
As before, we denote the scale where this occurs as $\Lambda_N$.  A vacuum
alignment argument implies that the symmetry breaking is such as to leave the
largest residual symmetry.  This implies that the condensate breaks the
original SU($N$) gauge symmetry to ${\rm SU}(N-1) \otimes {\rm U}(1)$.  Without
loss of generality, we take the breaking direction in SU($N$) to be $a=N$. To
show how this occurs, we recall the decompositions of $\sym_{\ {\rm SU}(N)}$
and $\overline{\asym}_{\ {\rm SU}(N)}$ under SU($N-1$) given, respectively, in
Eqs. (\ref{symdecomp}) and (\ref{asymdecomp}) above.  Using these
decompositions, we have
\begin{widetext}
\beq
\sym_{\ {\rm SU}(N)} \times \overline{\asym}_{\ {\rm SU}(N)} = 
\Big ( \sym_{\ {\rm SU}(N-1)} + \fund_{\ {\rm SU}(N-1)} + 1 \Big ) \times 
\Big (\overline{\asym}_{\ {\rm SU}(N-1)} + \overline{\fund}_{\ {\rm SU}(N-1)}
\Big ) \ . 
\label{sfdecomp}
\eeq
\end{widetext}
Among the various products, we see that 
$\fund_{\ {\rm SU}(N-1)} \times \overline{\fund}_{\ {\rm SU}(N-1)}$ yields
a singlet of SU($N-1$) and hence
is favored by the vacuum alignment argument.  The associated condensate thus 
has the form (with no sum on $a$) 
\beq
\langle T^a_a \rangle = \cases{ \kappa & for $1 \le a \le N-1$ \cr
                          -\kappa(N-1) & for $a=N$ } 
\label{taaform}
\eeq
where $\kappa$ is a constant. Thus, in terms of the fermion fields, the 
$\langle T^a_a \rangle$ condensate is of the form 
$\langle T^a_a \rangle = \langle \psi^{ad \ T}_L C 
\chi_{da,n_{\bar A},L}\rangle$ (with no sum on $a$ or $d$). 
The fermions involved in this condensate gain dynamical masses
of order $\Lambda_N$.  The $2N$ gauge bosons in the coset 
${\rm SU}(N)/[{\rm SU}(N-1) \otimes {\rm U}(1)]$ 
also gain masses of order $\Lambda_N$.  In the
low-energy effective field theory that is
applicable at scales $\mu < \Lambda_N$, one thus integrates out these fields
with masses $\sim \Lambda_N$. 

The massless nonsinglet chiral fermion content of the resultant 
low-energy effective field theory with 
${\rm SU}(N-1) \otimes {\rm U}(1)$ gauge invariance thus 
consists of the $\sym_{\ {\rm SU}(N-1)}$, $p$ copies of the
            $\overline{\asym}_{\ {\rm SU}(N-1)}$, and $p-1$ copies of the
 $\overline{\fund}_{\ {\rm SU}(N-1)}$, with corresponding U(1) charges. 
We summarize this SU($N-1$)-nonsinglet content as
\beq
{\rm SU}(N-1): \quad {\rm fermions}: \ S + p \, \bar A + (p-1) \bar F
\label{sunminus1fermions}
\eeq
where here $S$, $\bar A$, and $\bar F$ refer to the SU($N-1$) gauge
symmetry. This SU($N-1$) effective theory also contains the massless
SU($N-1$)-singlet fermion $\psi^{NN}_L$.  As guaranteed by the theorem proved
above, this descendant theory is free of anomalies in gauged currents.  
This is evident for the ${\rm SU}(N-1)^3$ triangle anomaly, for example, 
since this is given (with $M \equiv N-1$) by:
\beqs
{\rm SU}(N-1)^3 \ {\cal A} & = & 
{\cal A}(S)+p {\cal A}(\bar A) + (p-1){\cal A}(\bar F) 
\cr\cr
& = & (M+4)-p(M-4)-(p-1) = 0 \ , 
\cr\cr
& & 
\label{anomcancellation_sunminus1}
\eeqs
where the last line follows upon substitution of $p$ from Eq. (\ref{p}).
Similar cancellations hold for the ${\rm SU}(N-1)^2 \, {\rm U}(1)$ and 
U(1)$^3$ anomalies. 


\subsection{ SU(6) Theory with $n_S=1$, $n_{\bar A}=5$}

The renormalization-group evolution of a $(N;n_S,n_{\bar A})$ theory into the
infrared depends on the specific theory.  For definiteness, we shall focus on
the (6;1,5) theory for our further discussion.  We list the values of the first
two coefficients of the beta function for this theory in Table
\ref{minimal_andcopies}.  As before, since $n_S=1$ for this theory, we use a
simplified notation without the flavor index on the $S$ field, namely
$\psi^{ab}_{i=1,L} \equiv \psi^{ab}_L$. Our discussion for general $N \ge 6$
applies, in particular, to this theory.


\subsubsection{ Initial Condensation and Breaking of SU(6) to 
${\rm SU}(5) \otimes {\rm U}(1)$} 

Since the ratio $\rho$ is substantially larger than unity (see Table
\ref{minimal_andcopies}) and since we have not found composite fermion
operators that satisfy the 't Hooft anomaly matching conditions, we infer that
as the reference scale $\mu$ decreases from the UV to the IR, the gauge
interaction produces a bilinear fermion condensate in the $S \times \bar A \to
adj$ channel. Using the notation introduced above, this occurs at a scale
denoted $\Lambda_6$. By convention, we take the breaking direction as $a=6$ and
the copy (flavor) label of the $\bar A_2$ fermion involved to be $j=p=5$. In
the notation of Eq. (\ref{taaform}), the condensate can then be written as
\beq
\langle T^a_a \rangle = 
\langle \psi^{a6 \ T}_L \chi_{6a,5,L}\rangle
\label{su6tosu5condensate}
\eeq
where $1 \le a \le 5$, and there is no sum on $a$. The fermions involved in
this condensate gain dynamical masses of order $\Lambda_6$. 


\subsubsection{Analysis of Descendant ${\rm SU}(5) \otimes {\rm U}(1)$ Theory} 

We next consider the descendant ${\rm SU}(5) \otimes {\rm U}(1)$
theory that emerges from the self-breaking of the SU(6) theory at $\Lambda_6$.
The relevant decomposition of the SU(6) $S$ and $\bar A$ representations under 
${\rm SU}(5) \otimes {\rm U}(1)$, are indicated as follows, in terms of 
Young tableaux and SU(5) dimensionalities, with U(1) charges given as 
subscripts (normalized according 
to the conventions of \cite{slansky}):
\beqs
\sym_{\ {\rm SU}(6)} & = & [\ \sym + \fund + 1 \ ]_{\ {\rm SU}(5)} 
\cr\cr
                     & = & 15_{2} + 5_{-4} + 1_{-10} 
\label{symdecomp_su6tosu5}
\eeqs
and
\beqs
\overline{\asym}_{\ {\rm SU}(6)} & = &
[ \ \overline{\asym} + \overline{\fund} \ ]_{\ {\rm SU}(5)} \cr\cr
& = & \overline{10}_{-2} + \overline{5}_{4}                  
\label{asymdecomp_su6tosu5}
\eeqs
We will again use the shorthand notation $S \equiv S_2$, 
$\bar A \equiv \bar A_2$, 
and $\bar F$ for the $\sym $, $\overline{\asym}$ and $\bar{\fund}$, where
these now refer to SU(5).  We will indicate the U(1) charge of the $S$ field as
$\eta_S$ and so forth for the other fermion fields.  The massless
SU(5)-nonsinglet fermion content in this effective theory is thus
\beq
{\rm SU}(5): \quad {\rm fermions}: \ S \ + \ 5 \, \bar A \ + \ 4 \ \bar F \ . 
\label{su5descendant_fermions}
\eeq
Explicitly, these fermions (with dimensions of the SU(5) representations
indicated in parentheses) are 
\beqs
& & S (15): \quad \psi^{ab}_L \ {\rm with} \ 1 \le a, \ b \le 5, \cr\cr
& & 5 \ \bar A (10): \quad \chi_{ab,j,L} \ {\rm with} \ 1 \le a, \ b \le 5 \ 
{\rm and} \ 1 \le j \le 5, \cr\cr
& & 4 \ \bar F (5): \quad \chi_{6b,j,L} \ {\rm with} \ 1 \le b \le 5 \ 
{\rm and} \ 1 \le j \le 4 \ . \cr\cr
& & 
\label{su5leeft_explicitfermions}
\eeqs
This theory also contains the massless SU(5)-singlet fermion $\psi^{66}_L$ from
the original SU(6) theory. From Eq. (\ref{symdecomp_su6tosu5}), it follows that
this fermion has U(1) charge $\eta_1=-10$.

As an illustration of our theorem, it is instructive to see explicitly the
cancellation of contributions to the anomalies in various gauge currents in
this ${\rm SU}(5) \otimes {\rm U}(1)$ descendant gauge theory. 
There are three triangle anomalies that are relevant, namely the 
${\rm SU}(5)^3$, ${\rm SU}(5)^2 \, {\rm U}(1)$, and U(1)$^3$ anomalies.  We
have
\beqs
{\rm SU}(5)^3 \ {\cal A} & = & {\cal A}(S) + 5 \, {\cal A}(\bar A) + 
4 \, {\cal A}(\bar F) \cr\cr & = & 9 +5(-1)+4(-1) = 0
\label{su5cubed_anomaly}
\eeqs
and
\beqs
{\rm SU}(5)^2 \, {\rm U}(1) \ {\cal A} & = & T(S) \eta_S + 5 \, T(\bar A)
\eta_{\bar A} + 4 \, T(\bar F)\eta_{\bar F} \cr\cr
& = & \frac{7}{2} \, 2 + 5 \, \frac{3}{2}(-2) + 4 \, \frac{1}{2}(4) = 0 \ .
\cr\cr
& & 
\label{su5square_u1_anomaly}
\eeqs
For the ${\rm U}(1)^3$ anomaly cancellation, we must also include the
contribution of the SU(5)-singlet fermion $\psi^{66}_L$ since it carries a
nonzero U(1) charge: 
\beqs
& & {\rm U}(1)^3 \ {\cal A} = {\rm dim}(S)\eta_S^3 + 5 \, {\rm dim}(\bar A) 
\eta_{\bar A}^3 \cr\cr
& + & 4 \, {\rm dim}(\bar F) \eta_{\bar F}^3 + \eta_1^3 
\cr\cr
& = & 15(2^3) + 5(10)(-2)^3 + 4(5) (4^3) + (-10)^3  = 0 \ . \cr\cr
& & 
\label{u1cubed_anomaly}
\eeqs
We also observe that the mixed gauge-gravitational anomaly vanishes: 
\beqs
& & ({\rm grav})^2 \, U(1) \ {\cal A} = {\rm dim}(S)\eta_S + 
5 \, {\rm dim}(\bar A) \eta_{\bar A} \cr\cr
& + & 4 \, {\rm dim}(\bar F) \eta_{\bar F} + \eta_1 
 \cr\cr
& = & 15(2) + 5(10)(-2) + 4(5)(4) + (-10) = 0 \ . 
\cr\cr
& & 
\label{gravgravu1_anomaly}
\eeqs

The first two (reduced) coefficients of the SU(5) beta function are $\bar b_1 =
0.76925$ and $\bar b_2 = -0.22882$, so that this two-loop SU(5) beta function
has an IR zero at $\alpha_{IR,2\ell}=-\bar b_1/{\bar b_2} = 3.36$.  The U(1)
beta function is not asymptotically free, so that as the reference scale $\mu$
decreases, the running U(1) gauge coupling inherited from the original SU(6)
theory decreases. As regards the SU(5) dynamics, the most attractive channel
for fermion condensation is $S \times \bar F \to F$, with 
$\Delta C_2 = 28/5$. 
(The next-most attractive channel is $\bar A \times \bar A \to F$, with
$\Delta C_2 = 24/5$.)  Fermion condensation in this most attractive channel
causes the gauge symmetry breaking 
\beq
{\rm SU}(5) \otimes {\rm U}(1) \to {\rm SU}(4) \ . 
\label{su5xu1_to_u1}
\eeq
The fact that the fermion condensate breaks the U(1) gauge symmetry is evident,
since $\eta_S + \eta_{\bar F} = 6 \ne 0$.  The estimated minimum critical
coupling for condensation in this MAC to occur is $\alpha_{cr} \simeq 0.38$.
Since the ratio $\rho = \alpha_{IR,2\ell}/\alpha_{cr}=9.0$, and since we have
not found ${\rm SU}(5) \otimes {\rm U}(1)$-invariant fermionic operator
products that satisfy 't Hooft anomaly matching, we anticipate that fermion
condensation occurs in this most attractive channel. We denote the scale at
which this occurs as $\Lambda_5$. The $S \bar F$ condensate is of the form
$\langle \sum_{b=1}^5 \psi^{ab \ T}_L C \chi_{b6,j,L}\rangle$.  By convention,
we denote the breaking direction as $a=5$ and choose the copy index on the
$\chi_{b6,j,L}$ field to be $j=4$, so that the condensate is
\beq
\langle \sum_{b=1}^5 \psi^{5b \ T}_L C \chi_{b6,4,L}\rangle \ . 
\label{sfbar_condensate_su5tosu4}
\eeq
The fermions $\psi^{5b}_L$ and $\chi_{b6,4,L}$ with $1 \le b \le 5$ that are
involved in this condensate gain dynamical masses of order $\Lambda_5$. The ten
gauge bosons in the coset $[{\rm SU}(5) \otimes {\rm U}(1)]/{\rm SU}(4)$ 
also gain masses of order $\Lambda_5$. These fields are integrated out in the
construction of the SU(4)-invariant low-energy effective field theory 
applicable at scales $\mu < \Lambda_5$. 


\subsubsection{Analysis of Descendant ${\rm SU}(4)$ Theory} 

The massless SU(4)-nonsinglet chiral fermion content of this effective
low-energy theory consists of $\sym$, 5 copies of $\overline{\asym}$, and eight
copies of $\bar{\fund}$, i.e.
\beq
{\rm SU}(4): \quad {\rm fermions}: \ S + 5 \, \bar A + 8 \, \bar F
\label{su4leeft_fermions}
\eeq
Explicitly, these fermions (with dimensions of the SU(5) representations
indicated in parentheses) are 
\beqs
& & S (10): \quad \psi^{ab}_L \ {\rm with} \ 1 \le a, \ b \le 4, \cr\cr
& & 5 \ \bar A (6): \quad \chi_{ab,j,L} \ {\rm with} \ 1 \le a, \ b \le 4 \ 
{\rm and} \ 1 \le j \le 5, \cr\cr
& & 5 \ \bar F (4): \quad \chi_{5b,j,L} \ {\rm with} \ 1 \le b \le 4 \ 
{\rm and} \ 1 \le j \le 5 \cr\cr
& & 3 \ \bar F (4): \quad \chi_{6b,j,L} \ {\rm with} \ 1 \le b \le 4 \ 
{\rm and} \ 1 \le j \le 3 \cr\cr
& & 
\label{su4leeft_explicitfermions}
\eeqs
This theory also contains the
massless SU(4)-singlet fermions $\psi^{66}_L$ and $\chi_{65,j,L}$ with $1 \le j
\le 3$.

In accordance with our theorem, we show explicitly that this SU(4) 
descendant theory is anomaly-free: 
\beqs
{\rm SU}(4)^3 \ {\cal A} & = & {\cal A}(S) + 5 \, {\cal A}(\bar A) + 
8 \, {\cal A}(\bar F) \cr\cr & = & 8 + 0 + 8(-1) = 0 \ . 
\label{su4cubed_anomaly}
\eeqs
where we have used the fact that the $\overline{\asym} = \bar A$
representation of SU(4) is self-conjugate.

The first two (reduced) coefficients of the beta function of this SU(4)
descendant theory are $\bar b_1 = 0.5305$ and $\bar b_2=0.1224$, with the same
sign, so at the maximal scheme-independent, two-loop level, the beta function
has no IR zero.  Hence, as the reference scale decreases below $\Lambda_5$, the
SU(4) gauge coupling inherited from the SU(5) theory continues to increase,
eventually exceeding the range where it is perturbatively calculable.  The most
attractive channel for the formation of a bilinear fermion condensate is
\beq
\bar A \times \bar A \to 1 \ , 
\label{abarabarto1}
\eeq
with $\Delta C_2 = 2C_2(A) = 5$. Clearly, this fermion condensation preserves 
the SU(4) gauge symmetry.  The estimated minimal critical coupling for
condensation in this channel is $\alpha_{cr}=2\pi/15 = 0.42$. The associated
condensates are of the form 
$\langle \epsilon^{abde} \chi_{ab,j,L}^T C \chi_{de,k,L}\rangle$, where 
the copy indices take on values in the interval $1 \le j,k \le 5$.  Applying a 
vacuum alignment argument, we may take $j=k$, so that these condensates are
\beq
\langle \epsilon^{abde} \chi_{ab,j,L}^T C \chi_{de,j,L}\rangle \quad {\rm for}
\ 1 \le j \le 5
\label{aato1}
\eeq
(where there is no sum on $j$). These condensates are equal and hence preserve
an O(5) isospin symmetry.  We denote the scale at which this condensation takes
place as $\Lambda_{\bar A \bar A}$.  Owing to this condensation, all of the 
$\chi_{ab,j,L}$ fields gain masses of order  $\Lambda_{\bar A \bar A}$. 

This leaves a descendant (anomaly-free) chiral gauge theory with massless
SU(4)-nonsinglet fermion content $S + 8 \bar F$, given by
Eq. (\ref{su4leeft_explicitfermions}) with the $\bar A$ fields removed. This
theory has been studied before \cite{by,ads,cgt,net}, and we can combine the
known results with the new ingredients here for our analysis.  The first two
(reduced) coefficients in the beta function are $\bar b_1 = 0.7958$ and $\bar
b_2 = 0.2913$, with the same sign, so that this beta function has no IR zero.
Hence, as the scale $\mu$ decreases below $\Lambda_{\bar A \bar A}$, the SU(4)
gauge coupling continues to increase from its value at 
$\Lambda_{\bar A \bar A}$.

Since it is known that this $S + 8\bar F$ theory satisfies the 't Hooft anomaly
matching conditions \cite{by,ads,cgt}, one possibility is that it confines
without any spontaneous chiral symmetry breaking, producing massless composite
fermions and massive hadrons.  An alternate type of IR behavior is fermion
condensation in the most attractive channel, which is $S \times \bar F \to F$,
breaking SU(4) to SU(3), followed by further fermion in the respective MAC $S
\times \bar F \to F$ channels in the descendant SU(3) and SU(2) theories,
finally breaking the gauge symmetry completely.


\subsubsection{Discussion} 

It is of interest to contrast our present SU($N$) $S \bar A$ theories with 
$N \ge 6$, and hence a most attractive channel of the form 
$S \times \bar A \to adj$, 
with the theories analyzed in Ref. \cite{isb}.  One of the purposes of
Ref. \cite{isb} was to investigate how a fermion condensate transforming as the
adjoint representation of a simple SU($N$) gauge theory would dynamically 
break the gauge symmetry, and to contrast this with the types of gauge symmetry
breaking patterns that one obtains if one uses a fundamental Higgs field 
transforming according to the adjoint representation of the SU($N$) group.  
The type of theory considered in \cite{isb} had a direct-product gauge group 
of the form
\beq
G_{UV} = G \otimes G_b \ , 
\label{guv}
\eeq
where $G$ is a chiral gauge symmetry and $G_b$ is a vectorial gauge
symmetry. As constructed, the $G_b$ gauge interaction becomes strong in the
infrared and leads to a condensate involving a fermion field transforming as a
nonsinglet under both $G$ and $G_b$, and specifically as the adjoint
representation of $G$, thereby breaking $G$ to a subgroup, $H$.  At the stage
where this breaking occurs, the $G$ gauge interaction is still weak.  With
$G={\rm SU}(N)$ regarded as a hypothetical grand unification group,
Ref. \cite{isb} addressed the question of what the pattern of induced dynamical
breaking of a grand unified theory would be and how it would differ from the
pattern obtained with a nonzero vacuum expectation value of a fundamental Higgs
field in the adjoint representation.
This exploration of possible dynamical symmetry breaking of a grand unified
theory is reminiscent of, although different from, the old idea of dynamical
breaking of electroweak gauge symmetry by means of a vectorial, strongly
coupled, confining gauge theory which would produce bilinear fermion
condensates involving fermion(s) that transform under both the electroweak
gauge group and the strongly coupled gauge group \cite{weinberg76,tc}.  In the
latter case, the breaking of the electroweak gauge symmetry $G_{EW}$ is caused
by a bilinear fermion condensate transforming as the fundamental, rather than
adjoint, representation of weak SU(2)$_L$ with weak hypercharge 
$Y=1$.  

The difference with respect to our present work is that here we study a chiral
gauge theory with a single gauge group rather than a direct product, and the
chiral gauge interaction may produce condensates that self-break the strongly
coupled chiral gauge symmetry instead of having a weakly coupled chiral gauge
symmetry broken by a condensate of fermions that are nonsinglets under both $G$
and $G_b$.  The common feature shared by the dynamical gauge symmetry breaking
in studied in \cite{isb} and the $S \bar A$ theories with $N \ge 6$ here is
that the bilinear fermion condensate transforms as an adjoint of the SU($N$)
gauge symmetry and breaks it at the highest stage according to the pattern
(\ref{sabar_nge6_symmetrybreaking}).  To see how this differs with the
situation with a Higgs field $\Phi$ in the adjoint representation, we recall
the Higgs potential (with a $\Phi \to -\Phi$ symmetry imposed for technical
simplicity),
\beq
V = \frac{\mu^2}{2} {\rm Tr}(\Phi^2) +
\frac{\lambda_1}{4} [{\rm Tr}(\Phi^2)]^2 +
\frac{\lambda_2}{4} {\rm Tr}(\Phi^4) \ ,
\label{vsu5}
\eeq
where $\mu^2$, $\lambda_1$, and $\lambda_2$ are real for hermiticity. 
One chooses $\mu^2 < 0$ to produce the symmetry breaking.  Assuming
$N \ge 4$, for the comparison here, it follows that the two quartic terms in
(\ref{vsu5}) are independent, and the requirement that $V$ be bounded below
implies that $\lambda_1 > 0$.  This boundedness condition allows 
$\lambda_2$ to take on a restricted range of negative values 
depending on $\lambda_1$ and $N$, namely \cite{isb} 
\beq
-\bigg ( \frac{N(N-1)}{N^2-3N+3} \bigg ) \lambda_1 < \lambda_2 < 0 \ .
\label{lam2range}
\eeq
With $\lambda_2$ in this interval, $V$ is minimized with a Higgs VEV such that 
the SU($N$) gauge symmetry is broken according to 
(\ref{sabar_nge6_symmetrybreaking}).  However, if $\lambda_2 > 0$, then $V$ is
minimized for a Higgs VEV that yields a different symmetry breaking: if $N$ is
even, then the breaking pattern is
\beq
{\rm SU}(N) \to {\rm SU}(N/2) \otimes {\rm SU}(N/2) \otimes {\rm U}(1) \ , 
\label{symbreak_neven_lam2pos}
\eeq
while if $N$ is odd, then the breaking is 
\beq
{\rm SU}(N)\to {\rm SU}((N+1)/2) \otimes {\rm SU}((N-1)/2) \otimes
{\rm U}(1) \ .
\label{symbreak_nodd_lam2pos}
\eeq
This comparison elucidates the difference between the breaking of a gauge
symmetry by the VEV of a fundamental Higgs field and the dynamical symmetry
breaking by a fermion condensate produced by a strongly coupled gauge
interaction.


\section{Investigation of $S_k \bar A_k$ Chiral Gauge Theories with $k \ge 3$}
\label{kge3}

It is natural to ask whether the type of asymptotically free (anomaly-free)
chiral gauge theories that we have constructed and studied here with chiral
fermions transforming according to the rank-2 symmetric and conjugate
antisymmetric representations of SU($N$) can be extended to corresponding
chiral gauge theories with chiral fermions in the rank-$k$ symmetric and
rank-$\ell$ conjugate antisymmetric representations of SU($N$) with 
$k, \ \ell \ge 3$.  We show here that this cannot be done for the diagonal case
$k=\ell$ because such theories are not asymptotically
free.  Thus, our $S \bar A$ theories are the unique realization of 
asymptotically free $S_k \ \bar A_\ell$ chiral gauge theories with 
diagonal $k=\ell \ge 2$.

Let us then consider a chiral gauge theory with chiral fermions transforming
according to the rank-$k$ symmetric and conjugate antisymmetric representations
of SU($N$), denoted as the $S_k$ and $\bar A_k$. We denote the number of these
fermions as $n_{S_k}$ and $n_{\bar A_k}$, respectively, and the theory itself
as
\beq
(N;k;n_{S_k},n_{\bar A_k}) \ . 
\label{skakbar}
\eeq
The correspondence of this notation with the shorthand notation in the 
previous part of the text, which studied the $k=2$ case, is
\beq
(N;2;n_{S_2},n_{\bar A_2}) \equiv (N;n_S,n_{\bar A}) \ . 
\label{s2a2bar_correspondence}
\eeq
The condition that the theory must be free of any triangle anomaly in gauged
currents is
\beqs
& & n_{S_k} {\cal A}(S_k) + n_{\bar A_k} {\cal A}(\bar A_k) \cr\cr
& = & n_{S_k} {\cal A}(S_k) - n_{\bar A_k} {\cal A}(A_k) = 0 \ , 
\label{anomalycancellation_k}
\eeqs
where ${\cal A}(S_k)$ and ${\cal A}(A_k)$ are given in
Eqs. (\ref{anom_symk}) and (\ref{anom_asymk}) of Appendix
\ref{group_invariants}.  
A solution of this equation has the ratio of copies of fermions in the 
$S_k$ and $\bar A_k$ representations given by 
\beqs
\frac{n_{\bar A_k}}{n_{S_k}} & = & \frac{{\cal A}(S_k)}{{\cal A}(A_k)} 
\equiv p_k \cr\cr
 & = & \frac{(N+2k)(N+k)! (N-k-1)!}{(N-2k)(N+2)!(N-3)!} \ . 
\label{pk}
\eeqs
In the case $k=2$ discussed in detail above, $p_k = p$ given in Eq. (\ref{p}).
For $k \ge 3$, $p_k$ can also be expressed as 
\beq
p_k = \frac{(N+2k)\Big [\prod_{j=3}^k (N+j) \Big ]}
{(N-2k)\Big [\prod_{j=3}^k (N-j)\Big ]} \quad 
{\rm for} \ k \ge 3 \ . 
\label{pkkge3}
\eeq
For example, 
\beq
p_3 = \frac{(N+6)(N+3)}{(N-6)(N-3)}
\label{p3}
\eeq
and
\beq
p_4 = \frac{(N+8)(N+3)(N+4)}
           {(N-8)(N-3)(N-4)} \ . 
\label{p4}
\eeq

For a physical solution, this ratio (\ref{pk}) must be positive, which requires
that
\beq
N \ge 2k+1 \ , 
\label{nklowerbound}
\eeq
and we restrict $N$ to this range. From Eq. (\ref{pk}) it follows that if 
$k \ge 2$, then $n_{\bar A_k} > n_{S_k}$.  Therefore the theories of this 
type with minimal chiral fermion content have the form
\beq
(N;k;n_S,n_{\bar A}) = (N;k;1,p_k) \ , 
\label{nabark}
\eeq
with the understanding that $p_k$ must be a (positive) integer. If $k=3$, there
are only two solutions of Eq. (\ref{pk}) with $n_S=1$ that satisfy
this condition that $p_k$ be an integer, namely
\beq
(N;3;1,p_3) = \{ (9;3;1,10) \ , (12;3;1,5) \} \ . 
\label{minimalfermions_k3}
\eeq
The number of solutions decreases as $k$ increases.  Thus, if 
$k=4$, then there is only one such theory, viz.,
\beq
(N;4;1,p_4) = \{ (10;4;1,39) \} \ , 
\label{minimalfermions_k4}
\eeq
and similarly, if $k=5$, there is only one solution,
\beq
(N;5;1,p_5) = \{ (11;5;1,210) \} \ , 
\label{minimalfermions_k5}
\eeq
while we have not found solutions with integer $p_k$ for $k \ge 6$.  Theories
with $n_{cp}$ copies of this minimal fermion content also satisfy the anomaly
cancellation condition (\ref{anomalycancellation_k}), e.g., if $k=3$, then the
theories $(9;3;n_{cp},10n_{cp})$ and $(12;3;n_{cp},5n_{cp})$ for $n_{cp} \ge 2$
also satisfy the anomaly cancellation condition.

To test whether any of these solutions yield theories that are asymptotically
free, we begin by calculating the first coefficient of the beta function for
cases with minimal fermion content, with $n_{cp}=1$, which is 
\beqs
b_1 = \frac{1}{3}\Big [11N - 2( T_{S_k}+p_k T_{\bar A_k} ) \Big ] \ . 
\label{b1k}
\eeqs
If $k=3$, this is
\beqs
k=3 \ \rightarrow \ b_1 & = & 
\frac{1}{3}\bigg [ 11N - \frac{(N+3)(N^2-12)}{N-6} \bigg ] \cr\cr
& = & \frac{-N^3+8N^2-54N+36}{3(N-6)} \ . 
\label{b1k3}
\eeqs
Evaluating this for the two solutions (\ref{minimalfermions_k3}), we obtain
$\bar b_1=-4.695$ for $(N;k;1,p_3)=(9;3;1,10)$ and 
$\bar b_1=-5.252$ for $(N;k;1,p_3)=(12;3;1,5)$.  
These are both negative, so neither of these theories is asymptotically free. 

In a similar manner, we find that the anomaly-free $S_k \bar A_k$ theories with
higher $k$ are also not asymptotically free.  Substituting the value $k=4$ into
Eq. (\ref{b1k}) yields
\beqs
k=4 \ \rightarrow \ b_1 & = &
\frac{1}{3}\bigg [ 11N - \frac{(N+3)(N+4)^2(N-4)}{3(N-8)} \bigg ] \cr\cr
& = & \frac{-N^4-7N^3+37N^2-152N+192}{9(N-8)} \ . 
\cr\cr
& & 
\label{b1k4}
\eeqs
Evaluating this for the solution $(N;4;1,p_4)=(10;4;1,39)$, we obtain
$\bar b_1=-64.67$.  Finally, for $k=5$,
\beqs
& & k=5 \ \rightarrow \cr\cr
b_1 & = & 
\frac{1}{3}\bigg [ 11N - \frac{(N+3)(N+4)(N+5)(N^2-20)}{12(N-10)}\bigg ]\cr\cr
& = & \frac{-N^5-12N^4-27N^3+312N^2-380N+1200}{36(N-10)} \ . \cr\cr
& & 
\label{b1k5}
\eeqs
Evaluating this for the solution $(N;5;1,p_5)=(11;5;1,210)$, we get
$\bar b_1=-746.94$.  For each of these theories, letting $n_{cp}$ be
larger than 1 makes $b_1$ more negative, so the respective theories with
$n_{cp} \ge 2$ are also not asymptotically free.

Thus, we find that there are no anomaly-free chiral gauge theories with
fermions in the $k$-fold symmetric and conjugate antisymmetric representations
of SU($N$) for $k \ge 3$. 


\section{Investigation of $S_k \bar A_\ell$ Chiral 
Gauge Theories with $k \ne \ell$ and $k, \ \ell \ge 2$}
\label{skabarell}

One can also consider generalizations of our $S \bar A = S_2 \bar A_2$ chiral
gauge theories to theories with chiral fermions transforming as the rank-$k$
symmetric representation and the conjugate rank-$\ell$ antisymmetric
representation of SU($N$), where $k \ne \ell$ and $k, \ \ell \ge 2$.  We
consider theories of this type here. We denote the number of fermions
transforming as the rank-$k$ symmetric representation of SU($N$) as $n_{S_k}$
and the number of fermions transforming as the rank-$\ell$ conjugate
antisymmetric representation as $n_{\bar A_\ell}$, respectively, and the theory
itself as
\beq
(N;k;\ell;n_{S_k},n_{\bar A_\ell}) \ . 
\label{sk_abarell}
\eeq
Here the condition that there theory should have no anomaly in gauged currents
reads 
\beqs
& & n_{S_k}{\cal A}(S_k)+n_{\bar A_\ell} {\cal A}(\bar A_{\ell}) \cr\cr
& = & n_{S_k}{\cal A}(S_k) - n_{\bar A_\ell} {\cal A}(A_{\ell}) = 0 \ . 
\label{anomalycancellation_skabarell}
\eeqs
This anomaly cancellation condition is satisfied if and only if 
\beqs
\frac{n_{\bar A_\ell}}{n_{S_k}} & = & \frac{{\cal A}(S_k)}{{\cal A}(A_\ell)} 
 \equiv p_{\bar A_\ell/S_k}
\cr\cr
& = & \frac{(N+k)!(N+2k)(N-\ell-1)!(\ell-1)!}
           {(N-3)!(N-2\ell)(N+2)!(k-1)!}
\cr\cr
& & 
\label{p_nabarell_over_nsk}
\eeqs
Although $k \ne \ell$ here, we note that if one took $k=\ell$ as in the
previous part of this paper, then the correspondence in notation with
Eqs. (\ref{p}) and (\ref{pk}) is $p_{\bar A_k/S_k} \equiv p_k$ and $p_{\bar
  A_2/S_2} \equiv p_2 \equiv p$. For the ratio (\ref{p_nabarell_over_nsk}) 
to be a physical, positive number, it is necessary that
\beq
N \ge 2\ell+1 \ , 
\label{Nrangekell}
\eeq
and we shall restrict $N$ to this range. As explicit examples, we discuss
the $S_3 \bar A_2$ and $S_2 \bar A_3$ theories.


\section{$S_3 \bar A_2$ Theory}
\label{s3abar2}

Here, in accordance with (\ref{Nrangekell}), we restrict $N$ to the range 
$N \ge 5$. For this theory, the ratio $n_{\bar A_2}/n_{S_3}$ is 
\beq
\frac{n_{\bar A_2}}{n_{S_3}} \equiv p_{\bar A_2/S_3} = 
\frac{(N+3)(N+6)}{2(N-4)} \ . 
\label{ps3abar2}
\eeq
Unlike $p_k$ for the case of diagonal $S_k \bar A_k$ theories, 
this ratio $p_{\bar A_2/S_3}$ is
not a monotonic function of $N$.  It decreases from the value 44 at $N=5$ to a
formal minimum at the real value $N=4+\sqrt{70}=12.367$, where it is equal to 
$(17+2\sqrt{70})/2 = 16.667$, and then increases without bound as $N$ increases
further.  Since $p_{\bar A_2/S_3}$ is always larger than unity, it is natural
to consider models of this type with $n_{S_3}$ equal to its smallest value,
namely $n_{S_3}=1$.  We have found many of these, but none of them is
asymptotically free.  As allowed by the non-monotonicity of $p_{\bar A_2/S_3}$,
there are two values of $N$ that yield a minimal value of 
$p_{\bar A_2/S_3}$, namely $N=11$ and $N=14$, both of which give 
$p_{\bar A_2/S_3}=17$.  To minimize the fermion content with this value of 
$p_{\bar A_2/S_3}$, we choose $n_{S_3}=1$ and $n_{\bar A_2}=17$.  For $N=11$,
we have ${\cal A}(S_3)=119$ and ${\cal A}(\bar A_2)=-7$, while for $N=14$, we
have ${\cal A}(S_3)=170$ and ${\cal A}(\bar A_2)=-10$. To test whether any of
these solutions of the anomaly cancellation condition yields an asymptoticall
free theory, we calculate the one-loop coefficient of the beta function. In
general for this type of theory, 
\beq
b_1 = \frac{1}{3}\bigg [ 11N - 2 n_{S_3} \{T_{S_3} + 
p_{\bar A_2/S_3}T_{\bar A_2} \} \bigg ] \ . 
\label{b1s3abar2general}
\eeq
With the minimal choice $n_{S_3}=1$, this is
\beqs
b_1 & = & \frac{1}{3}\bigg [ 11N - \frac{(N+3)(N^2+N-10)}{N-4} \bigg ] \cr\cr
    & = & \frac{-N^3+7N^2-37N+30}{3(N-4)} \ . 
\label{b1s3abar2cases}
\eeqs
For the case with $N=11$, $\bar b_1= -3.263$, while for $N=14$, $\bar b_1 =
-4.934$.  These are both negative, i.e., these theories are not
asymptotically free.  Solutions of the anomaly conditions with larger values of
$n_{S_3}$ and $n_{\bar A_2}$ yield values of $b_1$ that are even more
negative.  Thus, we do not find any anomaly-free, asymptotically free 
theories of this $S_3 \bar A_2$ type. 


\section{$S_2 \bar A_3$ Theories} 
\label{s2abar3}

\subsection{General Analysis} 

Here we consider an SU($N$) theory with $n_{S_2}$ chiral fermions in the $S_2$
representation and $n_{\bar A_3}$ chiral fermions in the $\bar A_3$
representation. In accord with (\ref{Nrangekell}), we restrict $N$ to the range
$N \ge 7$. For this theory, the ratio $n_{\bar A_3}/n_{S_2}$ is
\beq
\frac{n_{\bar A_3}}{n_{S_2}} \equiv p_{\bar A_3/S_2} = 
\frac{2(N+4)}{(N-3)(N-6)} \ . 
\label{ps2abar3}
\eeq
This ratio decreases monotonically as a function of $N$ from the value value
11/2 at $N=7$ and approaches zero as $N \to \infty$.  The ratio
(\ref{ps2abar3}) takes on an integer value for only one value of $N$, namely
$N=10$, where it is equal to 1.  This reflects the equality ${\cal A}_{S_2} =
14 = {\cal A}_{A_3}$ for SU(10).  A theory with $N=10$ and $n_{cp} \ge 2$
copies of the $S_2$ and $\bar A_3$ representations is also anomaly-free.

For $N=10$ and $n_{S_2}=n_{\bar A_3}=1$, we calculate the reduced one-loop
coefficient in the beta function to be $\bar b_1=1.8569$, so this theory
satisfies the requirement of being asymptotically free.  We compute the
reduced two-loop coefficient to be $\bar b_2=0.086545$, so at the maximal
scheme-independent level, i.e., the two-loop level, this theory has no IR zero
in the beta function.  Hence, as the reference scale $\mu$ decreases from the
UV to the IR, the SU(10) gauge coupling continues to increase.

The condition of the cancellation of anomalies in gauged currents is also
satisfied in a theory in which the chiral fermion content is replicated
$n_{cp}$ times.  However, we find that only one of these nonminimal theories is
asymptotically free, namely the one with $n_{cp}=2$.  For this theory with
$N=10$ and $n_{S_2}=n_{\bar A_3}=n_{cp}=2$, we calculate $\bar b_1=0.795775$
and $\bar b_2=-7.00383$.  Thus, the two-loop beta function of this second
theory has an IR zero at $\alpha_{IR,2\ell}=0.1136$.


\subsection{SU(10) Theory with $n_{S_2}=n_{\bar A_3}=1$} 

\subsubsection{Initial Breaking of SU(10) to SU(6)} 

We will focus here on the simplest SU(10) theory of this type, with $n_{cp}=1$
and thus $n_{S_2}=n_{\bar A_3}=1$. This theory has a classical global symmetry
$G_{fl,cl}={\rm U}(1)_{S_2} \otimes {\rm U}(1)_{\bar A_3}$.  Both of these U(1)
symmetries are broken by SU(10) instantons, but one can construct a linear
combination U(1)$^\prime$ that is invariant in the presence of these
instantons.  Since (in a notation analogous to Eq. (\ref{anomvec})) 
${\vec v} =(T_{S_2},T_{\bar A_3}) = (6,14)$, U(1)$^\prime$ has the charge 
assignments
\beq
(Q_{S_2},Q_{\bar A_3}) \propto (7,-3) \ . 
\label{u1prime_s2abar3_charges}
\eeq
We have not found a set of gauge-singlet composite fermion operators 
satisfying the 't Hooft anomaly matching conditions for this U(1)$^\prime$
symmetry.  Therefore, we infer that as the SU(10) gauge coupling increases
sufficiently, fermion condensation will occur.  We find that the most
attractive channel is 
\beq
MAC: \quad \bar A_3 \times \bar A_3 \to A_4
\label{abar3abar3toa4channel}
\eeq
with 
\beq
\Delta C_2 = 9.90 \quad {\rm for} \ \bar A_3 \times\bar A_3 \to A_4 \ . 
\label{DeltaC2_abar3abar3toa4}
\eeq

We denote the $S_2$ and $\bar A_3$ fermion fields as $\psi^{ab}_L$ and
$\chi_{abcd,L}$.  The condensate for the channel $ \bar A_3 \times\bar A_3 \to
A_4$ is of the form
\beq
\langle \epsilon^{7 \, 8 \, 9 \, 10 \, \{a_1 ... a_6 \}}\chi^T_{a_1 a_2
  a_3,L}C\chi_{a_4 a_5 a_6,L}\rangle \ , 
\label{abar3abar3toa4condensate}
\eeq
where, by convention, we take the four uncontracted indices to be 7, 8, 9, and
10, and the summed indices to be $a_1,...,a_6 \in \{1,...,6\}$. We denote the
scale at which this condensate forms as $\Lambda_{10}$.  This condensate breaks
the SU(10) gauge symmetry to SU(6) and also breaks the global U(1)$^\prime$
symmetry. The 64 gauge bosons in the coset SU(10)/SU(6) also gain masses of
this order. In order to construct the low-energy effective SU(6) gauge theory
that is operative at reference scales $\mu < \Lambda_{10}$, we first enumerate
the chiral fermions that are involved in the condensate
(\ref{abar3abar3toa4condensate}) and that consequently gain dynamical masses of
order $\Lambda_{10}$ and are integrated out to form this low-energy effective
theory. The representation $\bar A_3$ has
dimension ${10 \choose 3} = 120$ in SU(10). One can choose the three
antisymmetrized group indices $a_1, \ a_2, \ a_3 \in \{1,...,6\}$ in the first
fermion in (\ref{abar3abar3toa4condensate}) in any of ${6 \choose 3} = 20$
ways, and the remaining group indices $a_4, \ a_5, \ a_6$ in any of 
${3 \choose 3}=1$ ways, so of the initial 120 components in the $\bar A_3$
fermion, the 20 components with gauge indices in the set $\{1,...,6\}$ gain 
masses and are integrated out of the SU(6) theory.

We next must determine how the remaining massless fermions transform under
SU(6).  For this purpose, let us use group indices $a,b,.. \in \{1,...,6\}$ to
refer to indices of the residual SU(6) gauge symmetry and $\alpha,\beta,.. \in
\{ 7, \ 8, \ 9, \ 10 \}$ to refer to the indices along the broken directions of
SU(10). The remaining 100 massless components of the $\bar A_3$ fermion can be
classified and enumerated as follows. First, there are the ${4 \choose 3}=4$
components $\chi_{\alpha \beta \gamma,L}$ for which 
$\alpha, \ \beta, \ \gamma \in \{7, \ 8, \, \ 9, \ 10\}$, 
which are singlets under SU(6).  Second, there
are the $6 \times {4 \choose 2} = 36$ components $\chi_{a \alpha \beta,L}$ with
$1 \le a \le 6$ and $7 \le \alpha, \ \beta \le 10$, which form six $\bar F$s of
SU(6). Third, there are ${6 \choose 2} \times 4 = 60$ components $\chi_{ab
 \alpha,L}$ with $1 \le a, \ b \le 6$ and $7 \le \alpha \le 10$, which comprise
four copies of $\bar A_2$ in SU(6).  For the
symmetric rank-2 tensor representation, we have
\beq
(S_2)_{{\rm SU}(10)} = (S_2)_{{\rm SU}(6)} + 4 \, F_{{\rm SU}(6)} + 
10 \, (1)_{{\rm SU}(6)}  \ . 
\label{s3decomp_su6fromsu10}
\eeq
where $(1)_{{\rm SU}(6)}$ is the singlet.  Recall that 
${\rm dim}(S_k) = (1/k!)\prod_{j=0}^{k-1}(N+j)$. 
Thus, the 55-dimensional $(S_2)_{{\rm SU}(10)}$ representation of SU(10) 
decomposes into the sum of the the $(S_2)_{{\rm SU}(6)}$
representation of SU(6) with its 21 component fields 
$\psi^{ab}_L$ with $1 \le a, b \le 6$,
plus four copies of the fundamental representation of SU(6) with fields
$\psi^{a \alpha}_L$, $1 \le a \le 6$ and $7 \le \alpha \le 10$, and ten
SU(6)-singlet fields $\psi^{\alpha \beta}_L$ with $7 \le \alpha, \ \beta \le
10$.  We summarize the massless SU(6)-nonsinglet chiral fermion content of the
low-energy SU(6) theory:
\beq
{\rm SU}(6): \quad {\rm fermions}: \ S_2 + 4 \bar A_2 + 4 \, F + 6 \, \bar F
\ . 
\label{su6_from_su10_fermions}
\eeq
The explicit fermion fields (with dimensionalities in parentheses) are
\beqs
& & S_2 (21): \quad \psi^{ab}_L \ {\rm with} \ 1 \le a, \ b \le 6, \cr\cr
& & 4 \ \bar A_2 (15): \quad \chi_{ab\alpha,L} \ {\rm with} 
\ 1 \le a, \ b \le 6 \ {\rm and} \ 7 \le \alpha \le 10, \cr\cr
& & 4 \ F (6): \quad \psi^{a\alpha,L} \ {\rm with} \ 1 \le a \le 6 \ {\rm and}
\ 7 \le \alpha \le 10, \cr\cr
& & 6 \ \bar F (6): \quad \chi_{a\alpha\beta,L} \ {\rm with} \ 1 \le a \le 6 \ 
{\rm and} \ 7 \le \alpha, \ \beta \le 10 \ . 
\cr\cr
& & 
\label{su6_from_su10__explicitfermions}
\eeqs
As guaranteed by our theorem above, this low-energy effective SU(6) theory
is anomaly-free; the contributions to the anomaly are
\beqs
{\cal A} & = & 
{\cal A}(S_2) - 4{\cal A}(A_2) +4{\cal A}(F) + 6{\cal A}(\bar F) \cr\cr
& = & 10 - (4 \times 2) + 4 - 6 = 0 \ . 
\label{anom_su6fromsu10}
\eeqs
%


\subsubsection{ Breaking of SU(6) to SU(5)}

We calculate the reduced one-loop and two-loop coefficients of the beta
function of this SU(6) theory (\ref{su6_from_su10_fermions}) to be 
$\bar b_1=0.84883$ and $\bar b_2=-0.56465$, 
so the two-loop beta function has an IR
zero at $\alpha_{IR,2\ell}=1.503$.  The most attractive channel for fermion
condensation is $S_2 \times \bar F \to F$, with $\Delta C_2 = 20/3$ and the
resultant estimate $\alpha_{cr} \simeq 0.31$. (The next-most attractive channel
is $F \times \bar F \to 1$ with $\Delta C_2 = 35/6$.) The ratio $\rho =
\alpha_{IR,2\ell}/\alpha_{cr} = 4.8$, which is considerably larger than
unity. We will explore evolution toward the
infrared that involves further fermion condensation, breaking 
the SU(6) gauge symmetry to SU(5). We denote the scale at which such
condensation occurs as
$\Lambda_6'$ (where the prime is included to avoid confusion with the scale
$\Lambda_6$ introduced in our discussion above of the SU(6) $S_2 \bar A_2$
theory). By convention, we label the breaking direction as $a=6$ and the
$\alpha,\beta$ indices of the $\bar F$ fermion as $\alpha=9, \ \beta=10$. The
associated $S_2 \bar F$ fermion condensate is then 
\beq
\langle \sum_{b=1}^6 \psi^{6b \ T}_L C \chi_{b\alpha\beta,L} \rangle \quad {\rm
  with} \ (\alpha,\beta)=(9,10) \ . 
\label{sfbarcondensate_su6}
\eeq 
The fermions involved in this condensate gain dynamical masses of order
$\Lambda_6$, as do the 11 gauge bosons in the coset SU(6)/SU(5).


\subsubsection{ Breaking of SU(5) to SU(4)}

To analyze the subsequent evolution into the infrared, we enumerate the
massless SU(5)-nonsinglet chiral fermion content of the resultant low-energy
effective SU(5) theory.  By the same methods as before, we find that this
content is
\beq
{\rm SU}(5): {\rm fermions}: \quad S_2 + 4 \bar A_2 + 4 F + 9 \bar F \ . 
\label{su5_from_su10_fermions}
\eeq
The explicit fermion fields (with dimensionalities in parentheses) are listed
below.  For this purpose, we relabel the group indices such that $a,b \in
\{1,...,5\}$ are SU(5) indices and $\alpha,\beta \in \{6,...,10\}$. We have
\beqs
& & S_2 (15): \quad \psi^{ab}_L \ {\rm with} \ 1 \le a, \ b \le 5, \cr\cr
& & 4 \ \bar A_2 (10): \quad \chi_{ab\alpha,L} \ {\rm with} 
\ 1 \le a, \ b \le 5 \ {\rm and} \ 7 \le \alpha \le 10, \cr\cr
& & 4 \ F (5): \quad \psi^{a\alpha}_L \ {\rm with} \ 1 \le a \le 5 \ {\rm and}
\ 7 \le \alpha \le 10, \cr\cr
& & 9 \ \bar F(5): \quad \chi_{a\alpha\beta,L} \ {\rm with} \ 1 \le a \le 5
\cr\cr
& & \quad\quad\quad\quad
  {\rm and} \ 6 \le \alpha, \ \beta \le 10 \ {\rm except } \ (\alpha,\beta) = 
(9,10) \cr\cr
& & 
\label{su5_from_su10__explicitfermions}
\eeqs

We calculate the reduced one-loop and two-loop coefficients of the beta
function of this SU(5) theory to be $\bar b_1=0.61009$ and $\bar b_2=-0.61384$,
so the two-loop beta function has an IR zero at $\alpha_{IR,2\ell}=0.994$.  The
most attractive channel for fermion condensation is $\bar A_2 \times \bar A_2
\to F$, with $\Delta C_2 = 24/5$ and the resultant estimate $\alpha_{cr} \simeq
0.44$.  The resultant ratio $\rho = \alpha_{IR,2\ell}/\alpha_{cr} = 2.3$,
suggesting that this condensation could plausibly occur. With condensation in
the $\bar A_2 \times \bar A_2 \to F$ channel, and with the breaking direction
taken to be $a=5$, the condensates are
\beqs
& & \langle \epsilon^{abcd5} \chi_{ab \alpha,L}^T C \chi_{cd \beta,L} \rangle 
\propto \Big [ \langle \chi_{12 \alpha,L}^T C \chi_{34 \beta,L}\rangle \cr\cr
& - & \langle \chi_{13 \alpha,L}^T C \chi_{24 \beta,L}\rangle + 
\langle \chi_{14 \alpha,L}^T C \chi_{23 \beta,L}\rangle \Big ] \ , 
\label{su5_from_su10_condensates}
\eeqs
where $6 \le \alpha, \ \beta \le 10$ as specified above. We denote the scale at
which these condensates form as $\Lambda_5'$. The fermions involved in these
condensates, as well as the nine gauge bosons in the coset SU(5)/SU(4), gain
masses of order $\Lambda_5'$.


\subsubsection{IR Evolution of the Descendant SU(4) Theory}

We determine the massless SU(4)-nonsinglet chiral fermion content of the
resultant SU(4) descendant theory to be
\beqs
{\rm SU}(4): & & \ {\rm fermions}: \ S_2 + 5 F + 13 \bar F \cr\cr
& = & S_2 + 8 \bar F + 5 \, \{F + \bar F\} \ . 
\label{su4_from_su10_fermions}
\eeqs
We see that this is precisely the $N=4$, $p=5$ special case of the $Sp$ model
of Eq. (\ref{spmodel}) studied in \cite{by,ads,cgt}.  For this theory we
calculate the beta function coefficients $\bar b_1 = 0.5303$ and $\bar b_2 =
-0.2496$, so the two-loop beta function has an IR zero at 
$\alpha_{IR,2\ell} = 2.125$.  
The fermion content of this theory satisfies the 't Hooft anomaly
matching conditions \cite{by,ads,cgt}, so one possibility is that as the gauge
interaction becomes strong, the theory confines and produces massless composite
SU(4)-singlet spin 1/2 fermions. Another possibility is that the gauge
interaction produces fermion condensation.  The most attractive channel is 
$S_2 \times \bar F \to F$ with $\Delta C_2 = 9/2$, so the rough estimate of
$\alpha_{cr}$ is $\alpha_{cr} \simeq 0.42$.  The resultant ratio $\rho=5.1$ is
well above unity, which renders it likely that either the gauge interaction
confines and produces the above-mentioned massless composite fermions or it
produces fermion condensation in this $S_2 \times \bar F \to F$ channel.  These
possibilities and the further evolution into the IR were discussed in detail in
\cite{cgt}. 


\section{Conclusions}
\label{conclusions}

In summary, in this paper we have constructed and studied asymptotically free
chiral gauge theories with an SU($N$) gauge group and $n_{S_k}$ copies of
massless chiral fermions transforming according to the symmetric rank-$k$
representation and $n_{\bar A_\ell}$ copies of fermions transforming according
to the conjugate antisymmetric rank-$\ell$ representation of this group, with
$k, \ \ell \ge 2$. As part of our work, we have proved a general theorem
guaranteeing that a low-energy effective theory resulting from the dynamical
breaking of an anomaly-free chiral gauge theory is also anomaly-free.  We have
explored the restrictions due to the constraints of asymptotic freedom and
anomaly cancellation and have shown that for a given $N$, $k$, and $\ell$,
these lead to, at most, a finite set of theories satisfying these restrictions.
For the case $k=\ell=2$, i.e., $S_2 \bar A_2$ chiral gauge theories, we have
given a detailed analysis of the UV to IR evolution of some simple theories,
including an SU(5) theory with $n_{S_2}=1$ and $n_{\bar A}=9$ and an SU(6)
model with $n_{S_2}=1$ and $n_{\bar A}=5$. We have shown that $S_2 \bar A_2$
theories exhibit a considerable variety of types of UV to IR evolution, ranging
from an infrared non-Abelian Coulomb phase to sequential chiral symmetry
breaking of both gauge and global chiral symmetry groups and possible
confinement with massless gauge-singlet composite fermions.  We have also shown
that there are no asymptotically free SU($N$) $S_k \bar A_k$ chiral gauge
theories with $k \ge 3$.  Finally, we have also studied chiral gauge theories
with chiral fermions in $S_k$ and $\bar A_\ell$ representations of SU($N$) with
$k \ne \ell$ and $k, \ \ell \ge 2$.  We believe that the results obtained here
give useful new insights concerning the properties of chiral gauge
theories. 


\begin{acknowledgments}
This research was partially supported by the NSF grant NSF-PHY-13-16617.
\end{acknowledgments}


\begin{appendix}

\section{Beta Function Coefficients and Relevant Group Invariants}
\label{beta_function}

For reference, we list the one-loop and two-loop coefficients
\cite{b1,b2} in the beta function (\ref{beta}) for a non-Abelian
chiral gauge theory with gauge group $G$ and a set of chiral fermions
comprised of $N_i$ fermions transforming according to the representations 
$R_i$: 
\beq
b_1 = \frac{1}{3}\Big [ 11 C_2(G) - 2 \sum_{R_i}N_i T(R_i) \Big ]
\label{b1gen}
\eeq
and
\beq
b_2=\frac{1}{3}\Big [ 34 C_2(G)^2 - 
2\sum_{R_i} N_i \{ 5C_2(G)+3C_2(R_i)\} T(R_i) \Big ] \ . 
\label{b2gen}
\eeq
%


\section{Relevant Group Invariants}
\label{group_invariants}

We list below the group invariants that we use for the relevant
case $G={\rm SU}(N)$.  The symmetric and antisymmetric rank-$k$ representations
of SU($N$) are denoted $S_k$ and $A_k \equiv [k]_N$. In terms of Young
tableaux, $S_1=A_1=\fund$, $S_2=\sym$, $A_2=\asym$, etc. (In the text, where no
confusion would result, we denote $S_2 \equiv S$ and $A_2 \equiv A$.)  For a
representation $R$, the Casimir invariants $C_2(R)$ and $T(R)$ are defined as
\beq
\sum_{i,j=1}^{{\rm dim}(R)} 
{\cal D}_R(T_a)_{ij} {\cal D}_R(T_b)_{ji}=T(R) \delta_{ab} 
\label{tr}
\eeq
and
\beq
\sum_{a=1}^{o(G)} \sum_{j=1}^{{\rm dim}(R)} 
{\cal D}_R(T_a)_{ij} {\cal D}_R(T_a)_{jk}=C_2(R)\delta_{ik} \ , 
\label{c2r}
\eeq
where $T_a$ are the generators of $G$, and ${\cal D}_R$ is the matrix
representation ({\it Darstellung}) of $R$.  These satisfy
\beq
T(R) \, o(G) = C_2(R) \, {\rm dim}(R) \ , 
\label{c2trelation}
\eeq
where $o(G)=N^2-1$ for SU($N$) and ${\rm dim}(R)$ is the dimension of the
representation $R$. 

For the adjoint representation, $C_2(adj) \equiv C_2(G) = T(Adj)=N$. 
For the rank-$k$ symmetric and antisymmetric representations $S_k$ and $A_k$, 
\beq
T(S_k) = \frac{\prod_{j=2}^k (N+j)}{2(k-1)!} 
\label{tr_symk}
\eeq
\beq
T(A_k) = \frac{1}{2}{N-2 \choose k-1} = \frac{\prod_{j=2}^k (N-j)}{2(k-1)!} 
\label{tr_asymk}
\eeq
\bigskip
\beq
C_2(S_k) = \frac{k(N+k)(N-1)}{2N}
\label{c2_symk}
\eeq
and
\beq
C_2(A_k) = \frac{k(N-k)(N+1)}{2N} \ . 
\label{c2_asymk}
\eeq
Hence, in particular, with $T_2$ standing for the rank-2 tensor representation
$S_2$ ($+$ sign) or $A_2$ ($-$ sign) here and below, one has 
\beq
T(T_2)=\frac{N \pm 2}{2}
\label{tr_rank2}
\eeq
\beq
C_2(T_2)=\frac{(N \pm 2)(N \mp 1)}{N}
\label{c2_rank2}
\eeq
\beq
T(T_3)=\frac{(N \pm 2)(N \pm 3)}{4}
\label{tr_rank3}
\eeq
and
\beq
C(T_3)= \frac{3(N \pm 3)(N \mp 1)}{2N} \ . 
\label{c2_rank3}
\eeq

The anomaly produced by chiral fermions transforming according to the 
representation $R$ of a group $G$ is defined as 
\beq
{\rm Tr}_R(T_a,\{T_b,T_c\}) = {\cal A}(R)d_{abc}
\label{anomgen}
\eeq
where the $d_{abc}$ are the totally symmetric structure 
constants of the corresponding Lie algebra.  Thus, 
${\cal A}(\fund)=1$ for SU($N$). For $S_k$ and $A_k$ \cite{anomtk}
\beq
{\cal A}(S_k) = \frac{(N+k)!(N+2k)}{(N+2)!(k-1)!}
\label{anom_symk}
\eeq
and
\beq
{\cal A}(A_k) = \frac{(N-3)!(N-2k)}{(N-k-1)! (k-1)!}  \ . 
\label{anom_asymk}
\eeq
Hence, in particular, 
\beq
{\cal A}(T_2) = N \pm 4
\label{anomaly_rank2}
\eeq
and
\beq
{\cal A}(T_3) = \frac{(N \pm 3)(N \pm 6)}{2} \ . 
\label{anomaly_rank3}
\eeq
%


\end{appendix} 


\newpage


\begin{widetext}

\begin{table}
\caption{\footnotesize{Properties of SU($N$) $S \bar A$ chiral gauge theories
    with (i) minimal fermion content $n_S=1$ and $n_{\bar A} = p=(N+4)/(N-4)$
    and (ii) $n_{cp}$-fold replicated fermion content $n_s=n_{cp}$ and $n_{\bar
      A}=n_{cp}p$. The quantities listed are $(N;n_S,n_{\bar A})$, $p$,
    $n_{cp}$, $\bar b_1$, $\bar b_2$, and, for negative $\bar b_2$,
    $\alpha_{_{IR,2\ell}}=-\bar b_1/\bar b_2$, $\alpha_{cr}$ for the relevant
    first condensation channel, and the ratio $\rho$ given by
    Eq. (\ref{rho}). The dash notation $-$ means that the two-loop beta
    function has no IR zero.  The likely IR behavior is indicated in the last
    column, where S$\chi$SB indicates spontaneously broken chiral symmetry,
    $\chi$S indicates a chirally symmetric behavior, and ESR stands for
    ``either symmetry realization'', $\chi$S or S$\chi$SB.  See text for 
    further discussion of descendant theories.}}
\begin{center}
\begin{tabular}{|c|c|c|c|c|c|c|c|c|}
\hline\hline
$(N;n_S,n_{\bar A})$ & $p$ & $n_{cp}$ & $\bar b_1$ & $\bar b_2$ & 
$\alpha_{_{IR,2\ell}}$ & $\alpha_{cr,ch}$ & $\rho_{_{IR}}$ & comment
\\ \hline
(5;1,9)  & 9 & 1 & 0.5570 & $-0.8638$ & 0.645  & 0.44 & 1.5   & ESR  \\
(6;1,5)  & 5 & 1 & 1.008  & $-0.1182$ & 8.53   & 0.39 & 22    & S$\chi$SB \\
(8;1,3)  & 3 & 1 & 1.592  & 0.9056    & $-$    & 0.28 & $-$   & S$\chi$SB \\
(12;1,2) & 2 & 1 & 2.600  & 3.519     & $-$    & 0.18 & $-$   & S$\chi$SB \\
\hline
(6;2,10) & 5 & 2 & 0.2653 & $-2.820$  & 0.0941 & 0.39 & 0.24  & $\chi$S \\
(8;2,6)  & 3 & 2 & 0.8488 & $-2.782$  & 0.3051 & 0.28 & 1.1   & ESR       \\
(12;2,4) & 2 & 2 & 1.698  & $-3.297$  & 0.5149 & 0.18 & 2.9   & S$\chi$SB \\
\hline
(8;3,9)  & 3 & 3 & 0.1061 & $-6.470$  & 0.0164 & 0.28 & 0.059 & $\chi$S   \\
(12;3,6) & 2 & 3 & 0.7958 & $-10.113$ & 0.07869& 0.18 & 0.44  & ESR \\
\hline\hline
\end{tabular}
\end{center}
\label{minimal_andcopies}
\end{table}

\newpage


\begin{table}
\caption{\footnotesize{Properties of SU($N$) $S \bar A$ chiral gauge theories
    with other $(N;n_S,n_{\bar A})$ than those in Table
    \ref{minimal_andcopies}. The quantities listed are $(N;n_S,n_{\bar A})$,
    $\bar b_1$, $\bar b_2$, and, for negative $\bar b_2$,
    $\alpha_{_{IR,2\ell}}=-\bar b_1/\bar b_2$, $\alpha_{cr}$ for the relevant
    first condensation channel, and the ratio $\rho$ given by
    Eq. (\ref{rho}). The dash notation $-$ means that the two-loop beta
    function has no IR zero.  The likely IR behavior is indicated in the last
    column, where S$\chi$SB indicates spontaneously broken chiral symmetry,
    $\chi$S indicates a chirally symmetric behavior, and $ESR$ stands for
    ``either symmetry realization'', $\chi$S or S$\chi$SB. See text for further
    discussion of descendant theories.}}
\begin{center}
\begin{tabular}{|c|c|c|c|c|c|c|}
\hline\hline
$(N;n_S,n_{\bar A})$ & $\bar b_1$ & $\bar b_2$ & $\alpha_{_{IR,2\ell}}$ & 
$\alpha_{cr,ch}$ & $\rho_{_{IR}}$ & comment
\\ \hline
(10;3,7) & 0.4775 & $-8.116$ & 0.0588  & 0.22 & 0.27 & $\chi$S \\
(16;3,5) & 1.379  & $-14.931$ & 0.0924 & 0.13 & 0.69 & ESR \\
(20;2,3) & 3.236  & $-4.265$  & 0.759  & 0.11 & 7.2& S$\chi$SB \\
(20;4,6) & 0.6366 & $-37.238$ & 0.0171 & 0.11 & 0.16 & $\chi$S \\
(28;3,4) & 3.024  & $-35.286$ & 0.0857 & 0.075& 1.1 & S$\chi$SB\\
(36;4,5) & 1.963  & $-102.512$& 0.0191 & 0.058& 0.33 & $\chi$S \\
(44;5,6) & 0.05305 & $-218.913$& 0.242e-3 & 0.048 & 0.0051 
& $\chi$S \\
\hline\hline
\end{tabular}
\end{center}
\label{other_theories}
\end{table}

\end{widetext}


\begin{thebibliography}{99}
\bibitem{sbara}
%
Clearly, our study of $S_k \bar A_\ell$ chiral gauge theories is
equivalent to the study of the set with fermions in the conjugate
representations $\bar S_k$ and $A_\ell$.
\bibitem{nur}
%
In the extended Standard Model that includes SM-singlet right-handed neutrinos,
there can be bare Majorana mass terms for these.
\end{thebibliography}
\end{document}